\documentclass{jaa}
\usepackage{natbib}
\usepackage{graphicx}
\usepackage{verbatim}
\usepackage[colorlinks, allcolors=blue]{hyperref}
\usepackage{adjustbox}
\usepackage{multirow}
\usepackage{multicol}
\usepackage[flushleft]{threeparttable}
\usepackage[utf8]{inputenc}

\begin{document}\sloppy

%%paper title
%%For line breaks \\ can be used within title
\title{UVIT study of UV bright stars in the globular cluster NGC 4147}

%%author names are separated by comma (,)
%%use \and before the last author name
%%use a * along with the number separated by comma
%% for the  author for correspondence
%%\textsuperscript{number} is used for affiliation
%%\affilOne, \affilTwo etc., upto \affilTwentyfive is possible
%%Please note the first letter after \affil is capitalised in the command

\author{Ranjan Kumar\textsuperscript{1*}, Ananta C. Pradhan\textsuperscript{1*}, Mudumba Parthasarathy\textsuperscript{2}, Devendra K. Ojha\textsuperscript{3}, Abhisek Mohapatra\textsuperscript{1}, Jayant Murthy\textsuperscript{2}, and Santi Cassisi\textsuperscript{4,5}}

\affilOne{\textsuperscript{1}Department of Physics and Astronomy, National Institute of Technology, Rourkela, Odisha - 769 008, India.\\}
\affilTwo{\textsuperscript{2}Indian Institute of Astrophysics, Koramangala II-Block, Bangalore - 560 034, India.\\}
\affilThree{\textsuperscript{3} Tata Institute of Fundamental Research (TIFR), Homi Bhabha Road, Mumbai - 400 005, India.\\}
\affilFour{\textsuperscript{4}INAF - Astronomical Observatory of Abruzzo, Via M. Maggini, sn. 64100 Teramo, Italy.\\}
\affilFive{\textsuperscript{5}INFN -  Sezione di Pisa, Universit\`a di Pisa, Largo Pontecorvo 3, 56127 Pisa, Italy.\\}
%%escape two column mode for title, affiliation and abstract
%%by giving \twocolumn command as shown

\twocolumn[{

\maketitle

%%include \corres to print the corresponding author Email id
\corres{ranjankmr488@gmail.com, acp.phy@gmail.com}

%%include \msinfo for
%%manuscript information such as
%%received, revised and accepted dates
%%
\msinfo{07 Nov 2020}{13 Dec 2020}

%%abstract
\begin{abstract}
We present far ultraviolet (FUV) observations of globular cluster NGC 4147 using three FUV filters, BaF2 (F154W), Sapphire (F169M), and Silica (F172M) of Ultra-Violet Imaging Telescope (UVIT) on-board the \mbox{{\em AstroSat}} satellite. We confirmed the cluster membership of the UVIT observed sources using proper motions from Gaia data release 2 (GAIA DR2). We identified 37 blue horizontal branch stars (BHBs), one blue straggler star (BSS) and 15 variable stars using UV-optical color magnitude diagrams (CMDs). We find that all the FUV bright BHBs are second generation population stars. Using UV-optical CMDs, we identify two sub-populations, BHB1 and BHB2, among the UV-bright BHBs in the cluster with stars count ratio of 24:13 for BHB1 and BHB2. The effective temperatures (T$_{\mathrm{eff}}$) of BHB1 and BHB2 were derived using color-temperature relation of BaSTI-IAC zero-age horizontal branch (ZAHB). We found that BHB1 stars are more centrally concentrated than BHB2 stars. We also derive physical parameters of the detected FUV bright BSS by fitting younger age BaSTI-IAC isochrones on optical and UV-optical CMDs. 
\end{abstract}

%%insert keywords separated by 3 hyphens using \keywords{words}
\keywords{(Galaxy:) globular clusters: individual: NGC 4147 - stars: horizontal-branch, (stars:) blue stragglers
- (stars:) Hertzsprung-Russell and colour-magnitude diagrams.}

}]

%%close the twocolumn escape here

%%include \doinum{number}for the DOI number in the header
%%include \volnum{number} for the volume number in the header
%%include \year{yyyy} for  year of publication in the header
%%include \pgrange{num--num} page range of article in the header
%%include \artcitid{num} for the article citation id
%%include \lp to print last page of the article
%%include \setcounter{page}{pagenum} for the exact starting page of the article

\doinum{12.3456/s78910-011-012-3}
\artcitid{\#\#\#\#}
\volnum{000}
\year{0000}
\pgrange{1--}
\setcounter{page}{1}
\lp{1}

\section{Introduction}
The glimpse of ultraviolet (UV) light in the old stellar population of the Galactic globular clusters (GGCs) is dominated by hot luminous UV-bright stars which are mostly the stars of blue horizontal branch (BHB), blue-stragglers (BS), post-asymptotic giant branch (pAGB), and extreme horizontal branch (EHB) phases having temperature more than 7,000 K. Various physical properties of these UV-bright populations have been explored using large sample of GGCs observed by {\em Galaxy Evolution Explorer} \citep[{\em GALEX},][]{Schiavon2012} and {\em Hubble space telescope} \citep[{\em HST},][]{Nardiello2018}. Ultraviolet Imaging Telescope (UVIT) on-board the \mbox{{\em AstroSat}} satellite \citep{Kumar2012} has also performed the imaging observations of several GGCs in UV with a better spatial resolution than {\em GALEX} enabling in resolving the core of the clusters and also in distinguishing UV bright stars of different evolutionary stages to study their physical parameters in a great detail \citep{Subramaniam2017, Sahu2019a, Jain2019, Kumar2020a, Kumar2020, Rani2020, Singh2020}. 
Using the spectroscopic data of a large number of stars in many GGCs along with the UV photometric observations with {\em HST}, the evidence for the presence of multiple stellar populations in GGCs is now well established \citep[see reviews][and references therein]{Bastian2018,Gratton2019, Cassisi2020}. The different chemical evolution within the GGCs seems to be the origin of two distinct sub-population of stars (the first generation (1G) and second generation (2G)) characterized by significant abundance variations in light elements. The stellar populations enriched in He, N, and depleted in O and C are 2G stellar population while stellar populations enriched in C and O and depleted in N and Na with a primordial He-abundance are termed as normal or 1G stellar population \citep[see ][and references therein]{Milone2017, Marino2019}.

We present here the UV study of an old age, metal poor, low density GC NGC 4147, located at a distance of 21 kpc from the Galactic center and 19 kpc from the Sun ($l = 252.84^\circ$, $b = +77.18^\circ$). We have observed this cluster using three far-UV (FUV) filters of UVIT. Prior to this, the UV imaging and color-magnitude diagrams (CMDs) of the cluster were presented by \cite{Schiavon2012} using {\em GALEX} observations. Their study was restricted to the sources in the outer region of the cluster due to the lower resolution ($\sim5''$) of {\em GALEX}. However, the cluster is well studied  both photometrically and spectroscopically in the optical bands of the electromagnetic spectrum. Several of the photometric observations in the optical bands have explored about specific sources of the cluster such as BHBs, variable stars, red giant branch stars (RGBs), etc., using their optical CMDs \citep{Auriere1991,Arellano2004, Stetson2005, Arellano2018, Lata2019}. Similarly, many low resolution spectroscopic observations of the cluster have been performed to find its metallicity ($[Fe/H]= -1.85$ dex), and chemical abundances of various alpha-elements \citep{Suntzeff1988, Martell2008, Ivans2009}. The latest high-resolution spectroscopy of 18 RGB stars by \cite{Villanova2016} has revealed that most of the RGB stars of the cluster are of 2G population type with 2G to 1G RGB stars' ratio of 85:15. They found that the red-HB stars (RHBs) are progeny of the 1G population and the BHBs are progeny of the 2G population. They also confirmed an alpha-enhancement of 0.38 dex and a mono-metallicity of $[Fe/H] = -1.84$ dex among 1G and 2G stars in the cluster. 

In \autoref{sec:data-reduction}, we present the observation details, data reduction process and photometry of the UVIT observation of the cluster. In \autoref{sec:cmd}, we show the CMDs of all the detected sources. In \autoref{sec:sub-population}, we discuss the sub-population among the UVIT detected 2G BHBs. In \autoref{sec:BSs}, we estimate various physical properties of the UV bright blue straggler star (BSS). Finally, we summarize our results in \autoref{sec:conclusion}.

\section{Data Reduction}
\label{sec:data-reduction}
We obtained the UVIT observation of the cluster NGC 4147 in three FUV filters, BaF2, Sapphire and Silica, having wide, medium and narrow bandwidths, respectively. The UVIT data of the cluster was obtained from \mbox{{\em AstroSat}} archival web-page\footnote{https://astrobrowse.issdc.gov.in/astro\_archive/archive/Home.jsp} in level1 format. The instrument and calibration details of UVIT can be found in \citet{Tandon2017, Tandon2020}. We reduced the level1 data into science image using a customized software package CCDLAB \citep{Postma2017}, specially designed for the UVIT data reduction. The UVIT observations have been performed in several orbits and we get science image for each orbit. In order to get a good signal to noise ratio (SNR), we combined all the observed orbits into a single image and then performed the photometry. The UVIT observation details are given in \autoref{tab:observation}. 

\begin{table}
\centering
\caption{Observational details of NGC 4147.}
\label{tab:observation}
\resizebox{\columnwidth}{!}{
\begin{tabular}{cccccc}
\hline
 \multicolumn{2}{c}{Date of observation} & \multicolumn{4}{l}{ 2017, April 17} \\
 \multicolumn{2}{c}{Telescope pointing} &  \multicolumn{4}{c}{ RA $=182.5263^\circ$,\hspace{10pt} 
 DEC $=18.5426^\circ$}\\
 \hline \hline
Channel & Filters & mean $\lambda$ & $\Delta \lambda$ &  Exp. Time & No. of Orbits \\
  & & (\AA) & (\AA) &  (sec.) \\
 \hline
\multirow{ 3}{*}{FUV} & BaF2 & 1541 & 380 & 1536 & 5  \\
                      & Sapphire &  1608 & 290 & 1648 & 3  \\
                      & Silica & 1717 & 125 & 1209 & 2 \\
\hline
\end{tabular}
}
\end{table}

\begin{figure}
    \centering
    \includegraphics[width=\columnwidth]{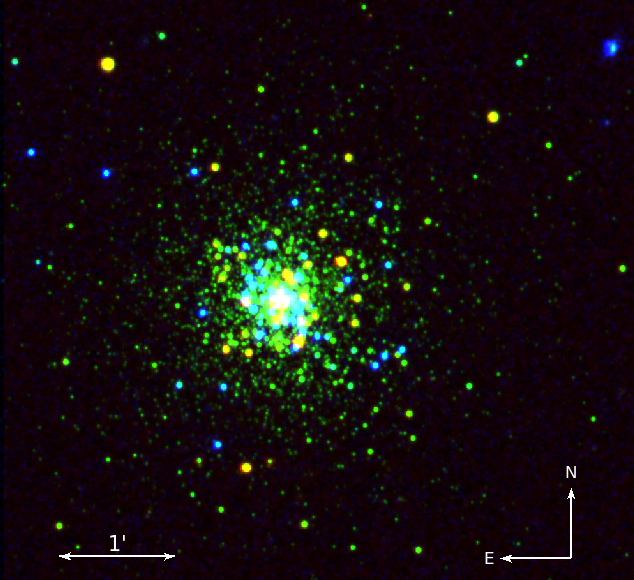}
    \caption{Color image of NGC 4147; UV: UVIT BaF2 (blue), Visible: CFHT-4m V band (green), Infrared: 2MASS J band (red). }
    \label{fig:color_image}
\end{figure}

In \autoref{fig:color_image}, we provide a color image of the cluster using multi-wavelength observations from infrared (IR) to UV. The sources in blue color are observed in UV using the UVIT BaF2 filter, the green color sources are from the archival catalog of CFHT-4m $V$ band observation \citep{Stetson2019}, and the red color sources are obtained from {\em 2MASS} $J$ band observation\footnote{https://irsa.ipac.caltech.edu/applications/2MASS} of the cluster. We can see that the hot sources, observed in UV (blue sources) are bright enough and easily distinguishable in the outer region of the cluster. The central region of the cluster also contains many hot sources as visible in cyan color (mixture of blue and green) due to crowding effect at the center.  

We applied DAOPHOT point spread function (PSF) photometry \citep{Stetson1987} on the science images to obtain the source positions and their respective apparent magnitudes. The aperture and saturation corrections were applied on the detected sources in all the three FUV filters following the suggestion by \cite{Tandon2017}. A typical PSF of 1.5$''$ was obtained in all the science images of the FUV filters. We have excluded the sources within 10$''$ from the cluster center to avoid contamination. We detected 114, 92, and 65 sources in the BaF2, Sapphire, and Silica filters, respectively. The AB-magnitude limits of the detected sources in BaF2, Sapphire and Silica filters are 23.0, 22.5, and 21.0, respectively. The magnitudes were corrected for extinction value using $E(B-V)=0.0221$ mag \citep{Schlafly2011} along the cluster direction and extinction law of \cite{cardeli1989} in the UV filters.

\section{Color-Magnitude Diagrams}
\label{sec:cmd}

\begin{figure*}
    \centering
    \includegraphics[width=0.495\textwidth]{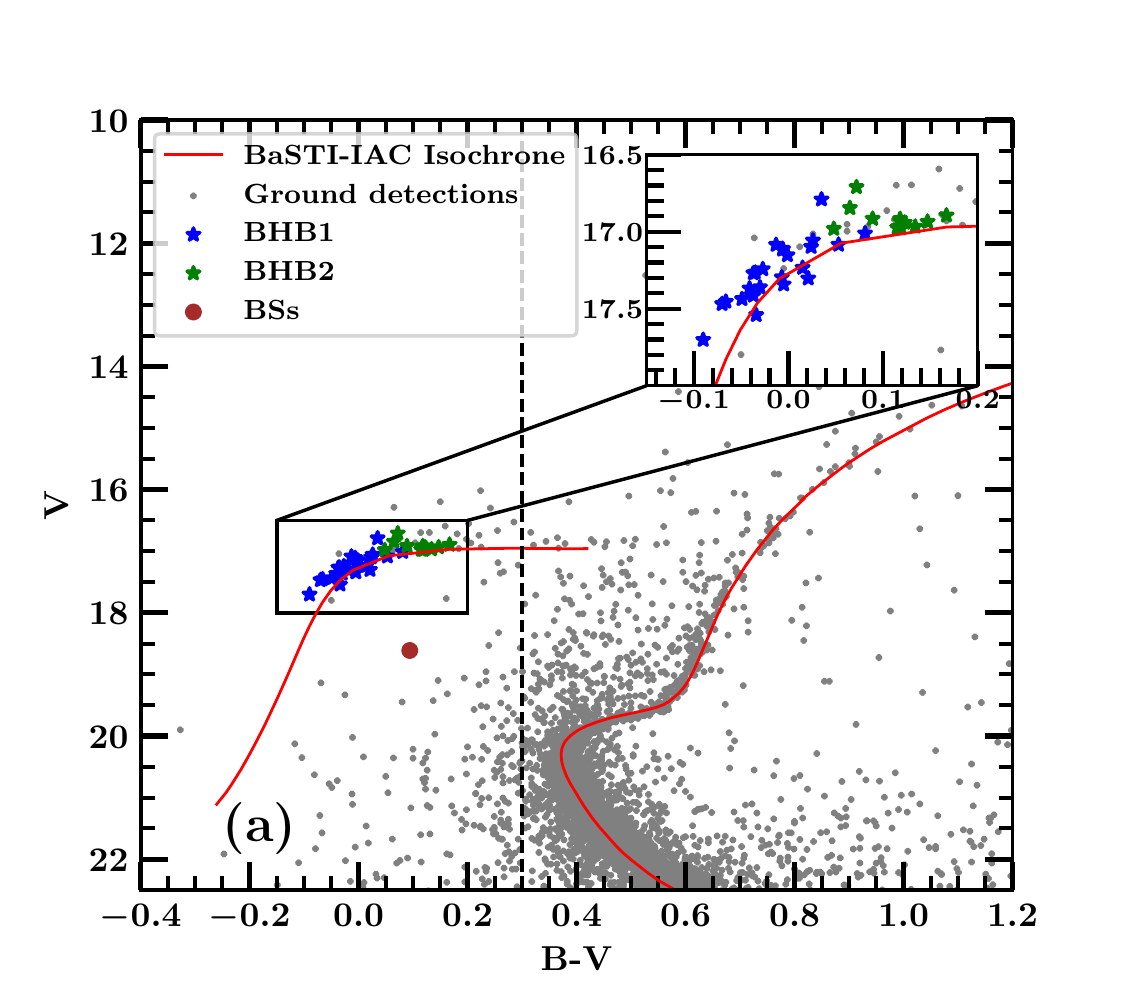}
    \includegraphics[width=0.495\textwidth]{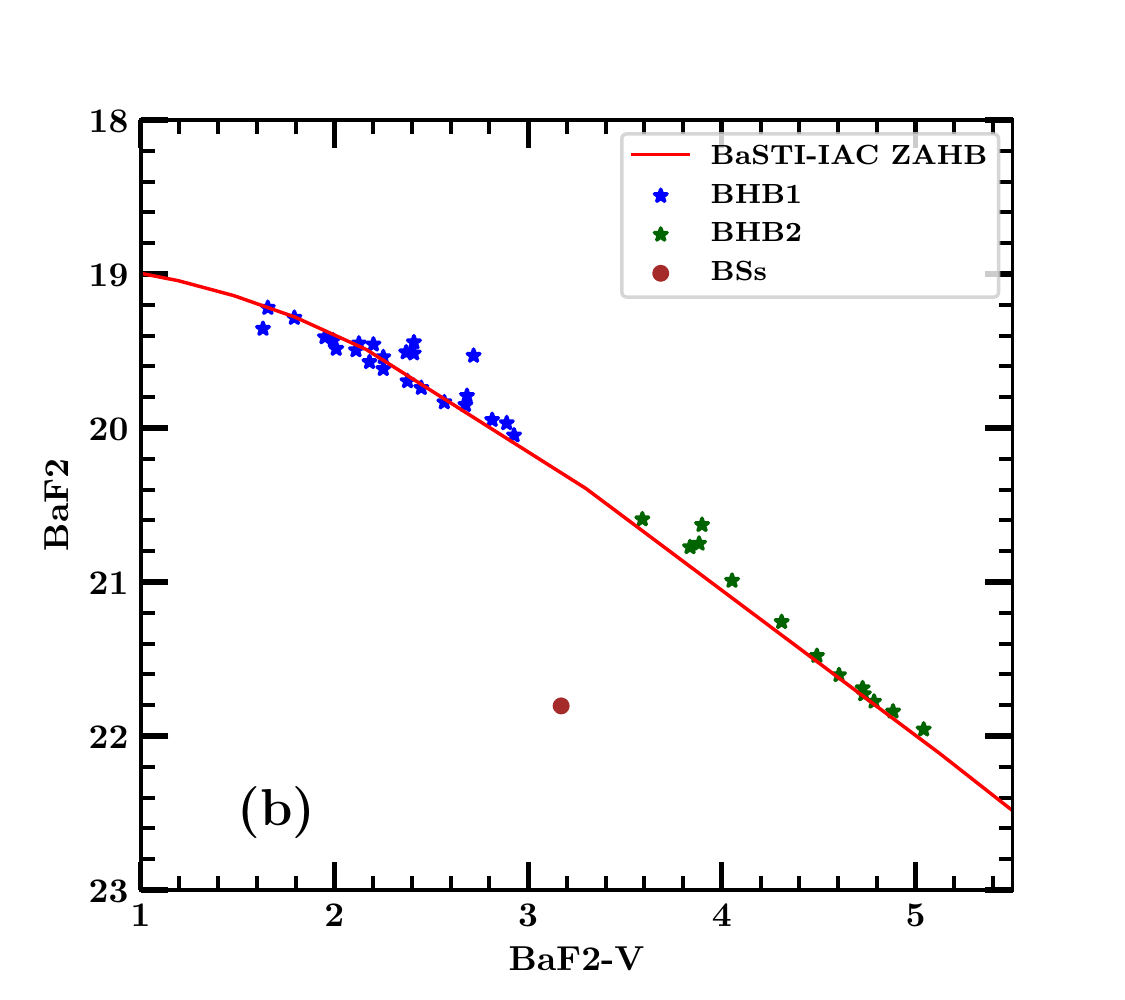}\\
    \includegraphics[width=0.495\textwidth]{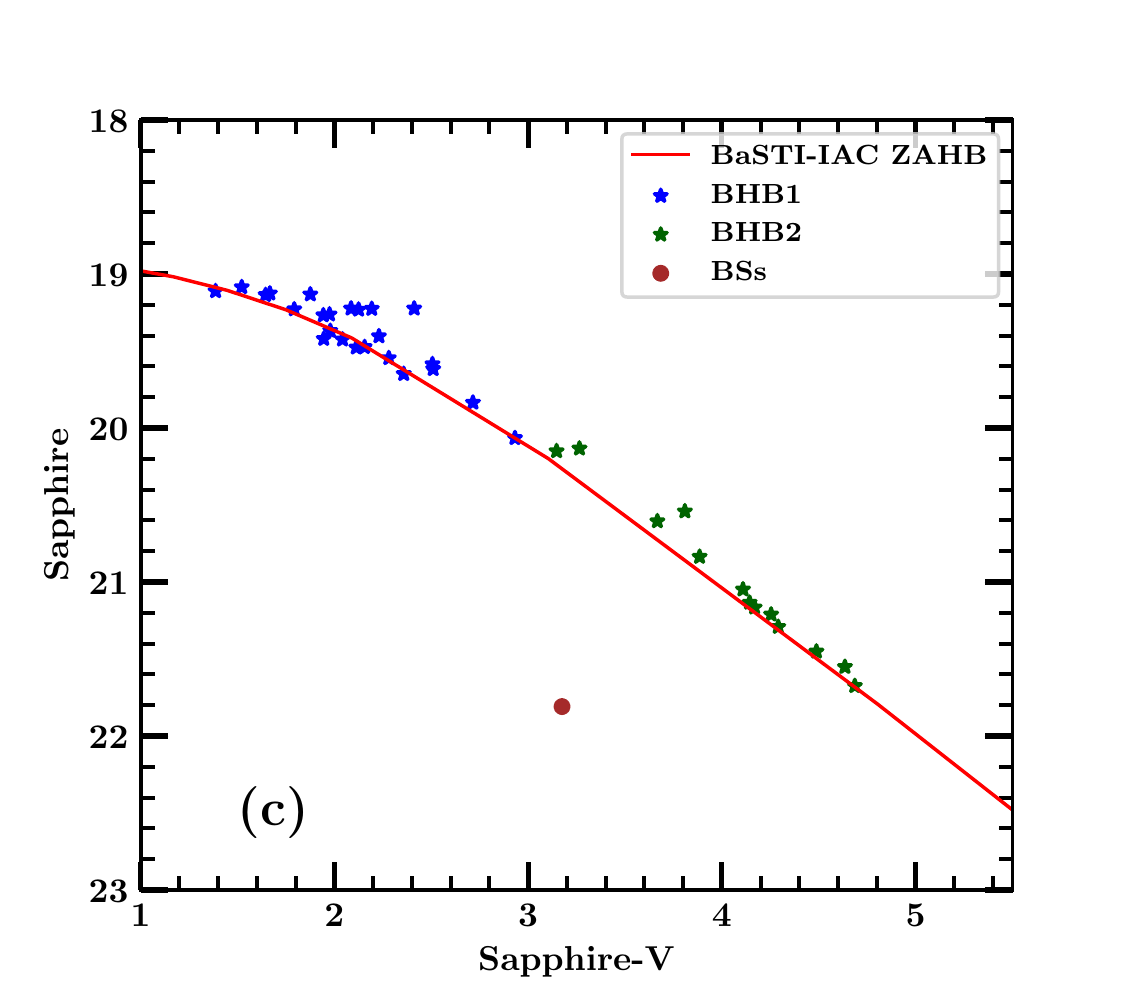}
    \includegraphics[width=0.495\textwidth]{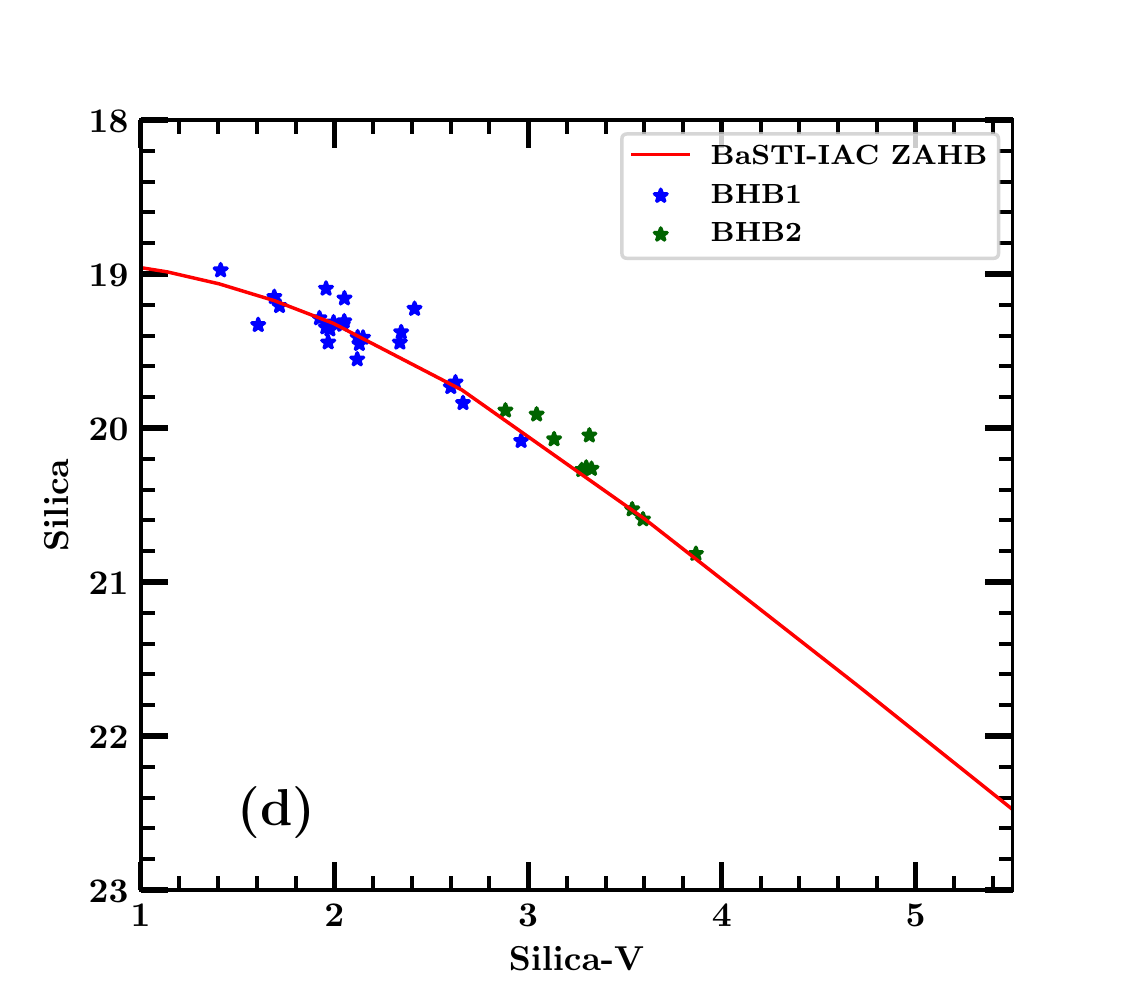}
    
    \caption{Optical and UV-optical CMDs of the UVIT observed sources of NGC 4147. \textbf{Panel (a)}: The B$-$V vs V CMD where ground-based detections from \cite{Stetson2019} are shown in gray solids, the UVIT observed BHBs are shown in blue and green asterisks. The BSS is shown in the brown solid. BaSTI-IAC isochrones and ZAHB with cluster parameter, $[Fe/H]=-1.896$, $[\alpha/Fe]=0.40$, Age = 13.0 Gyr, and distance-modulus $(m-M) = 16.40$, are over-plotted on the observed sources.\textbf{Panels (b), (c), and (d)}:  BaF2$-$V vs BaF2, Sapphire$-$V vs Sapphire, and Silica$-$V vs Silica CMDs, respectively, are shown for the UVIT observed sources with the same colors as mentioned in panel (a), and BaSTI-IAC ZAHB with the same cluster parameters as mentioned in panel (a) are over-plotted on all CMDs. }
    \label{fig:cmd}
\end{figure*}

To study UV and UV-optical CMDs of the cluster members, we cross-matched the UVIT observed sources with GAIA data release 2 catalog \citep[Gaia DR2,][]{GaiaCatalog2018} with a matching radius of 1.5$''$. The confirmed cluster members were separated out from the field stars and the background sources (Galaxies, quasars, etc.) using Gaia DR2 proper motions. Once the sources were confirmed as cluster members, we cross-matched them to deeper photometric observations, the latest released archival catalog of ground-based observations in UBVRI filters \citep{Stetson2019}. An optical CMD, B$-$V vs V, is shown in panel (a) of \autoref{fig:cmd} for all the ground-based observed sources within the UVIT field of view (FoV $\sim30'$) in gray solids. The UVIT detections are over-plotted in the CMD and are shown in blue and green asterisks and in brown solid circle.

In order to obtain the cluster properties, we used the Bag of Stellar Tracks and Isochrones models \citep[BaSTI-IAC\footnote{http://basti-iac.oa-abruzzo.inaf.it/index.html},][]{Hidalgo2018} and generated isochrones and zero-age horizontal branch (ZAHB) with cluster parameters, $[Fe/H]=-1.896$, age = 13.0 Gyr \citep{Harris2010} and $[\alpha/Fe]=0.40$ \citep{Villanova2016}. Since the alpha-enhanced set of the BaSTI-IAC library has not been still published, we relied on the extension of the BaSTI-IAC library to alpha-enhanced mixture \citep{Pietrinferni2020} for the present analysis. In panel (a) of \autoref{fig:cmd}, BaSTI-IAC isochrones are over-plotted on the B$-$V vs V CMD in red line using a distance-modulus of 16.40, which are fitting very well with the ground-based observations. In Panels (b), (c), and (d) of \autoref{fig:cmd}, we have shown the UV-optical CMDs of the cluster using BaF2, Sapphire, and Silica filters, respectively, in combination with V magnitude from the optical bands. The ZAHB from BaSTI-IAC library is over plotted on all the UV-optical CMDs. Although all the UV-optical CMDs look similar we have shown them to see the morphological distribution of UV bright stars of the cluster in all the FUV filters. We can see in \autoref{fig:cmd} that BHBs are major contributors to the UV bright sources of the cluster.

If we look at the optical CMD in panel (a) of \autoref{fig:cmd}, we notice that the UVIT observed BHBs have spread along the horizontal line of the HB phase. This spread is better visible in UV-optical CMDs along the diagonal HB line than the optical CMD.
In fact, the spread among BHBs in BaF2$-$V vs BaF2 CMD (panel (b)) becomes more clear with a gap of 0.8 magnitude in the BaF2$-$V color within BHBs. Based upon the gap observed in BaF2$-$V vs BaF2 CMD, we divide the UVIT observed BHBs in two groups: BHBs with BaF2$-$V color brighter than 2.8 magnitude as BHB1 (blue asterisks) and BHBs with BaF2$-$V color fainter than 3.6 magnitude as BHB2 (green asterisks). These BHBs are lying in left and right portions of the BHB region in B$-$V vs V CMD (panel (a)) and they are clearly identified in Sapphire$-$V vs Sapphire and Silica-V vs Silica CMDs in panels (c) and (d), respectively. There are 24 sources in BHB1 and 13 sources in BHB2 groups.

\cite{Villanova2016} have made a chemical abundance analysis of 18 RGB stars of the cluster NGC 4147 and found that the cluster is composed of 1G stars with $[Na/Fe]\sim +0.0$ and  $[O/Fe]\sim +0.3$ and 2G stars with $[Na/Fe]\sim +0.5$ and  $[O/Fe]\sim -0.2$. In Figure 1 of their paper they have shown that the 2G HB stars are located at $B-V<0.3$. If we compare the UVIT observations with optical data, it appears that all the observed BHBs seem to belong to the 2G population (panel (a) of \autoref{fig:cmd}). In the same figure, one can notice that the BHBs are separated in two sub-groups, which could point to the presence of two sub-populations in the observed 2G BHBs. We discuss this possibility in the following sections.

Apart from the BHBs, we have also detected one UV bright BSS in the cluster. The BSS was observed in BaF2 and Sapphire filters but not in the Silica filter. The brightness of the BBS may be fainter than the detection limit of Silica filter (21.0 AB-magnitude). The BSS is shown in \autoref{fig:cmd} with brown solid. We also found 15 variable stars by cross-matching the UVIT observed sources with the updated variable stars catalog of GCs \citep{Clement2017}. However, we have not shown these variable stars on the CMDs due to the severe crowding conditions affecting these stars, and the fact that we do not have a complete sampling of their light curves.

\section{Sub-population in BHBs}
\label{sec:sub-population}

\begin{figure}
    \centering
    \includegraphics[width=\columnwidth]{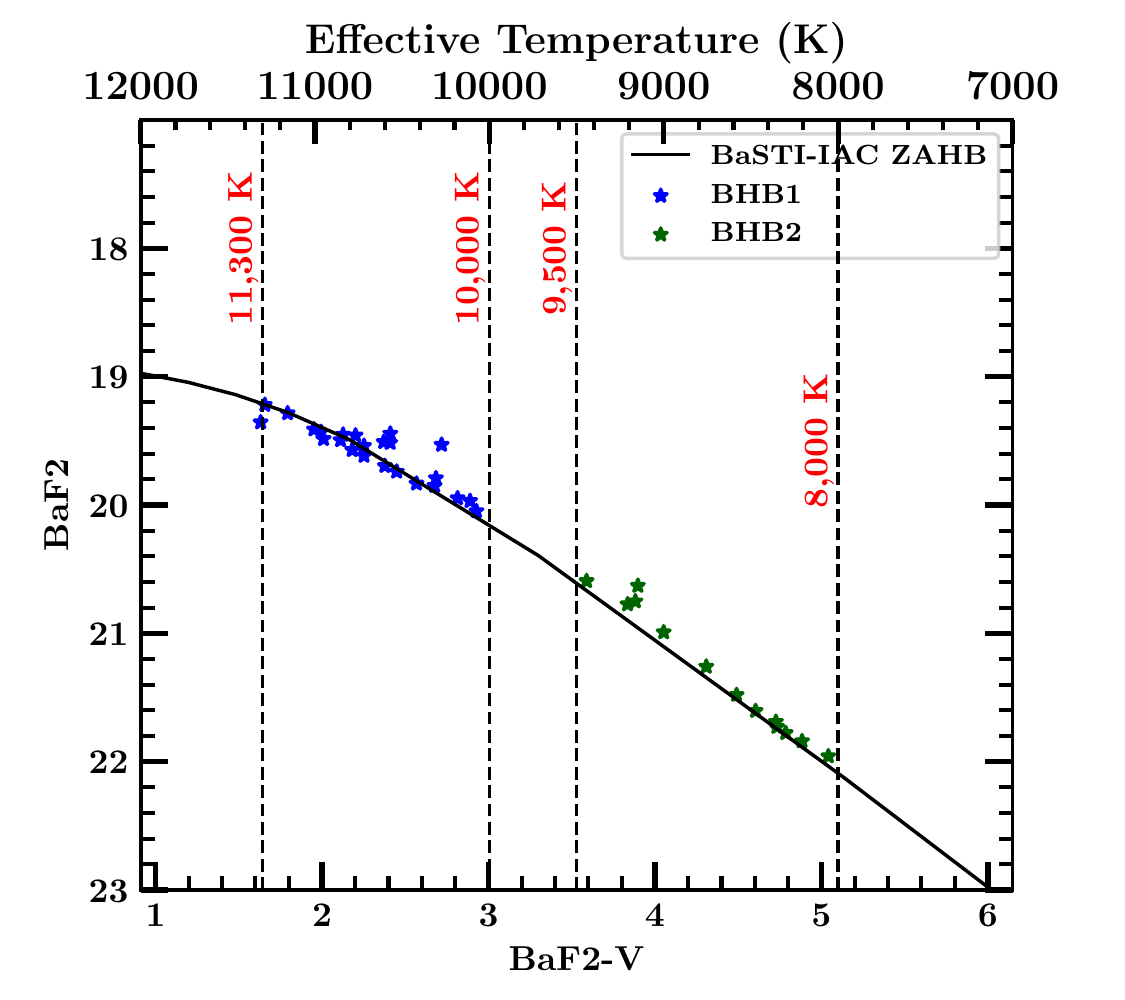}
    \caption{BaF2$-$V vs BaF2 CMD of the observed BHBs. We use BaSTI-IAC ZAHB color-temperature relation to derive the range of T$_{\mathrm{eff}}$ of BHB1 and BHB2. The T$_{\mathrm{eff}}$ derived from the ZAHB is given in the upper x-axis of the plot.  }
    \label{fig:color_teff}
\end{figure}

We derive the effective temperature (T$_{\mathrm{eff}}$) of the observed BHBs using color-temperature relation of the BaSTI-IAC ZAHB in \autoref{fig:color_teff}. The best fitted ZAHB from \autoref{fig:cmd} is taken to derive the color-temperature relation of the UVIT observed BHBs. As shown in \autoref{fig:color_teff}, the T$_{\mathrm{eff}}$ of BHB1 is ranging between 10,000 K and 11,300 K and T$_{\mathrm{eff}}$ of BHB2 ranges between 8,000 K and 9,500 K. The comparison between observations and models suggests the presence of a gap between BHB1 and BHB2 stars with a temperature difference of about 500 K. 
However, we caution that this width of the observed temperature gap and its actual existence may be affected by the limited number of sample stars (number of BHBs detected) in this cluster. Although a detailed statistical analysis of the reality of such a gap will be carried out in a future work, we note that a puzzling gap or discontinuity in the distribution of BHBs is seen in the color-magnitude diagram (CMD) of many GGCs \citep{Catelan1998,Piotto1999,Behr2000,Brown2016}. It was initially believed to be due to statistically significant under-population of stars but its appearance at similar location for different clusters has substantiated the claim that it is indeed a real feature and the satisfactory explanation for its origin is not yet known. However, speculation is that the gaps may be demarcating the boundaries between separate, discrete populations of HB stars, which differ in their origin or evolution.

By relying on the BaF2$-$V vs BaF2 CMDs obtained in the present work, we collect some evidence suggesting the possible presence of a bi-modality among the BHB population. In the following, we provide some suggestions about the possible origin for the observed discontinuity in the color distribution of BHB stars in NGC 4147. The main empirical findings are that: i) the BHB1 and BHB2 are clustered in the optical plane in two quite clear separate regions - although in that plane these two regions appear quite adjacent (panel (a) of \autoref{fig:cmd}); ii) in the UV-optical plane there exists an
evident gap of about 500 K whose width has been evaluated by over-imposing the models. These empirical evidence suggests that BHB1 stars could be slightly He enhanced with respect to BHB2 ones and in such a case they would represent two distinct sub-populations in the 2G population; or alternatively they could share with BHB1 more or less the same He abundance, and have experienced different mass loss efficiency during the previous RGB stage \citep[see, e.g., ][]{Tailo2020}.
 
\begin{figure}
    \centering
    \includegraphics[width=\columnwidth]{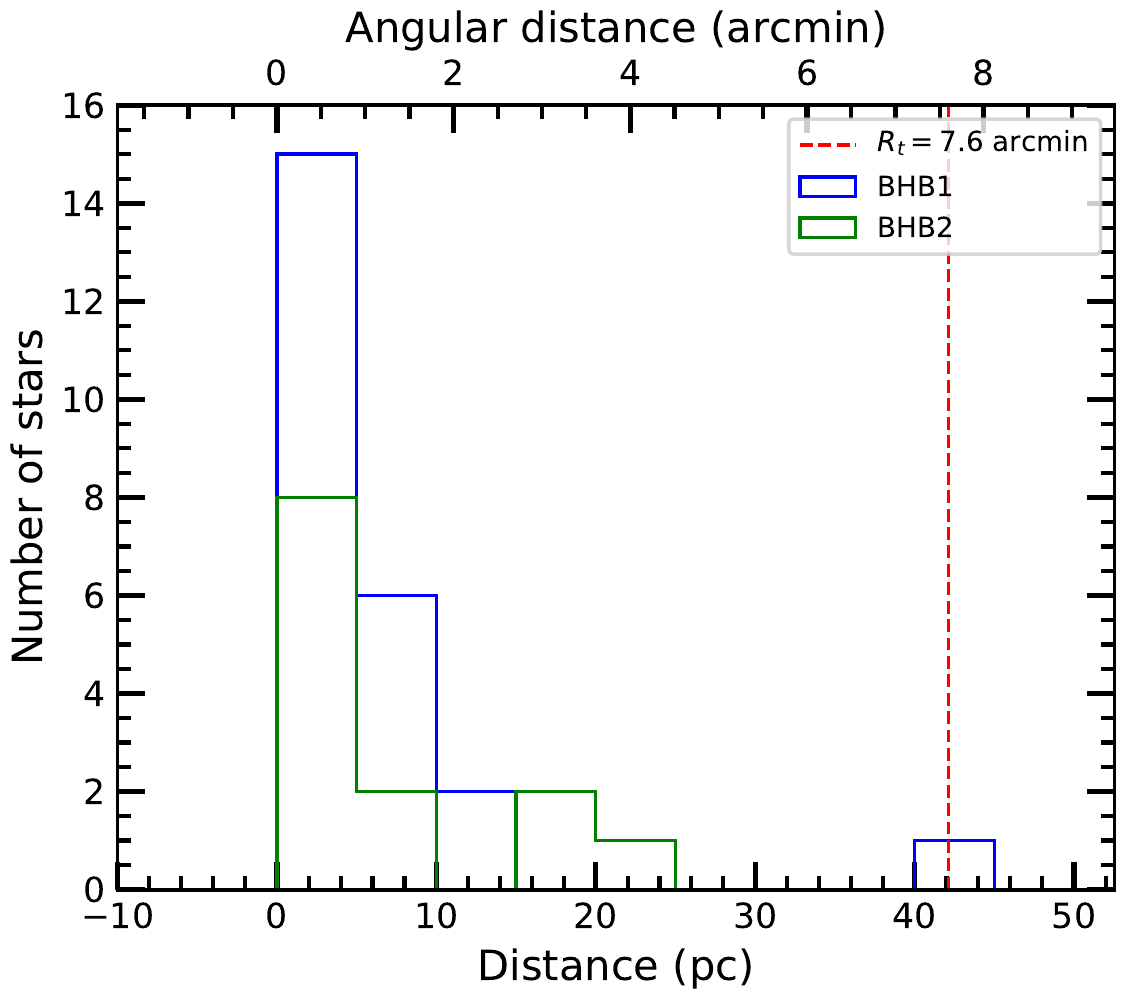}
    \caption{The radial distribution of UVIT detected BHBs. The histogram of BHB1 stars is shown in blue line and histogram of BHB2 stars is shown in green line. The lower and upper x-axes is the distance from the cluster center in parsec and arcminutes, respectively. The y-axes repesents the number of stars present in each distance bin of 5 parsec from the cluster center. The vertical dashed line is the tidal radius of the cluster.}
    \label{fig:bhb_radial}
\end{figure}

 In \autoref{fig:bhb_radial}, we have shown the radial distribution of BHB1 and BHB2 stars. We see that the BHB1 stars are spread within 15 pc (with an exception of one BHB1 star at a distance of 43 pc) from the cluster center. However, BHB2 stars show their radial distribution up to 25 pc. We find that the BHB1 stars are having almost double counts than the BHB2 stars up to 10 pc. This indicates that the hotter sources are relatively concentrated at the center than the lower T$_{\mathrm{eff}}$ BHBs.

\section{FUV bright Blue Straggler Star}
\label{sec:BSs}

The blue-stragglers (BS) are generally formed in the old age clusters through two formation channels: mass-transfer in binary system in low density environment \citep{Knigge2009,Leigh2013} and merger due to collision between stars in dense environment \citep{Chatterjee2013}. They are relatively more massive hydrogen burning MS stars than the normal MS population of the cluster owning to the higher mass and consequently shifted in the upper MS branch in the H-R diagram. The UV photometry is a great tool to identify the BS population in a cluster. The data from UV telescopes ({\em HST}, {\em GALEX} and UVIT) have been excellent tool to explore BS population and their formation scenario \citep{Ferraro1997, Ferraro2003, Dieball2010, Schiavon2012, Gosnell2015, Subramaniam2017, Sahu2019a, Kumar2020a, Kumar2020, Rani2020}. In particular, UVIT observations of star clusters are able to identify the hot binary companions (WD, HB, EHB, etc.) of BS stars \citep{Subramaniam2016, Sindhu2019, Sahu2019b, Jadhav2019, Singh2020}.

\begin{figure*}
    \centering
    \includegraphics[width=0.495\textwidth]{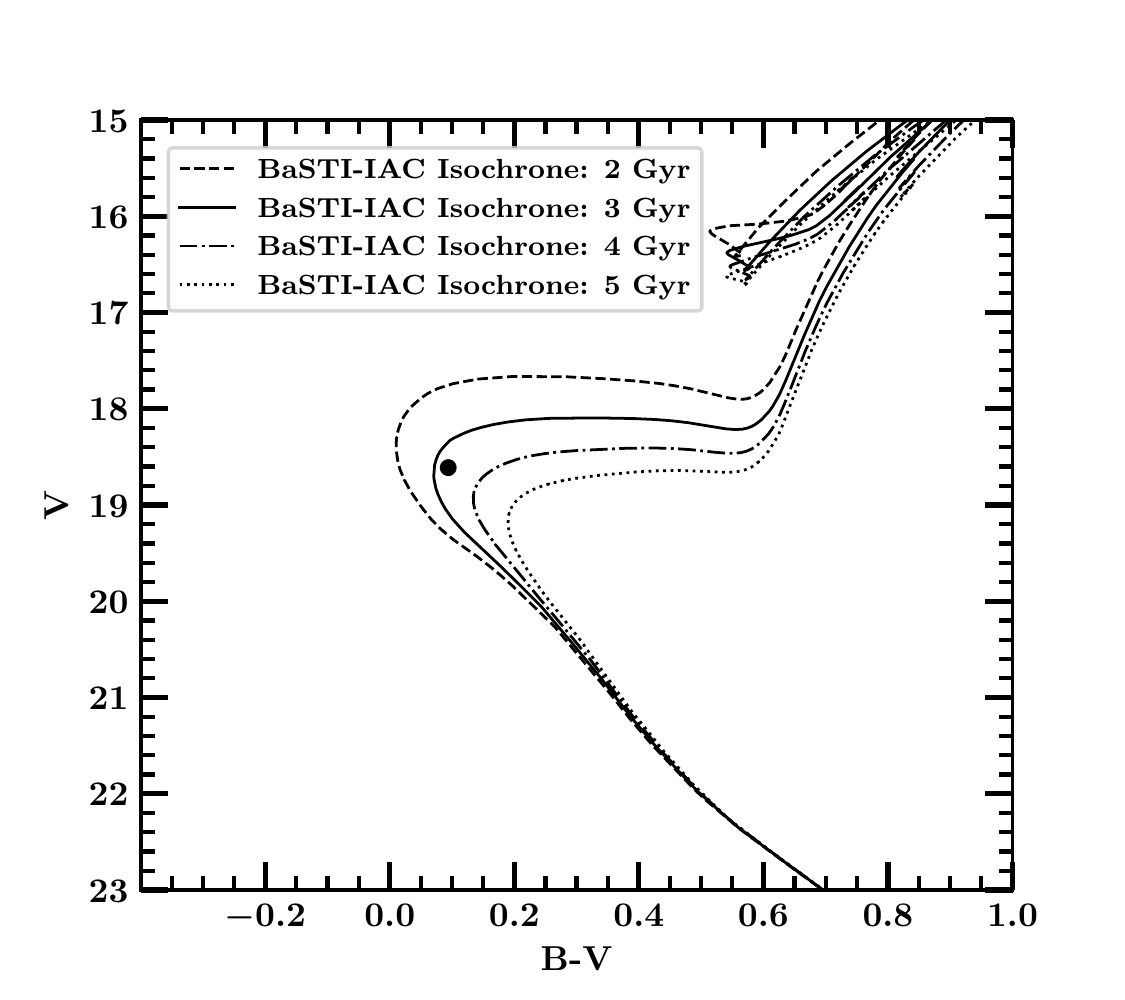}
    \includegraphics[width=0.495\textwidth]{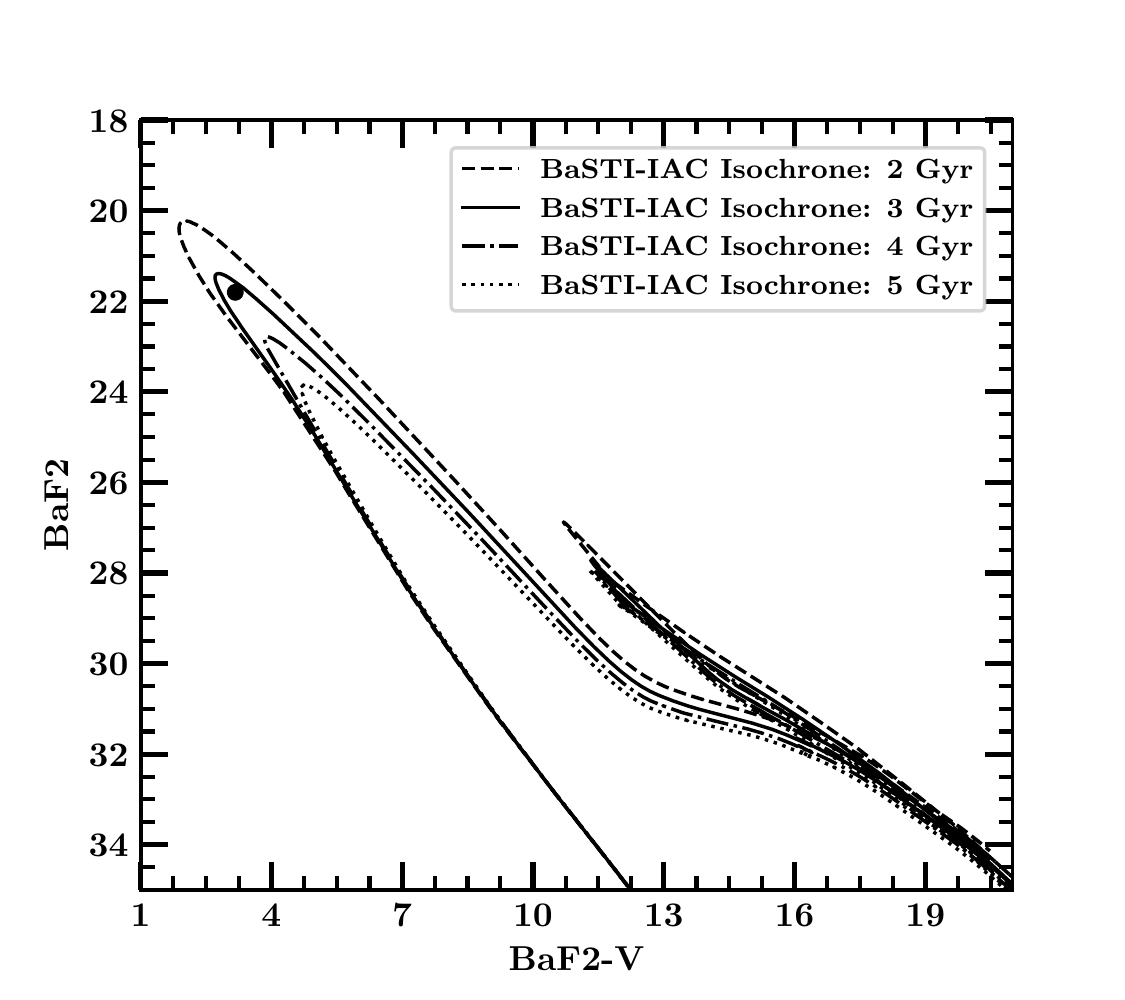}
    \caption{Optical and UV-optical CMDs are shown in left and right panels, respectively. The FUV bright BSS of NGC 4147 is shown in the black solid in both CMDs. The BaSTI-IAC isochrones with the same parameters as mentioned in \autoref{fig:cmd} but with younger ages, i.e 2 Gyr (dashed line), 3 Gyr (solid line), 4 Gyr (dash-dotted line), and 5 Gyr (dotted line), are over-plotted in both CMDs.}
    \label{fig:cmd_bs}
\end{figure*}

Present observational data-set allows us to identify one FUV bright BSS in the outer region of the cluster, at a distance of 2.15$'$ from the center. The respective colors and magnitudes of the BSS in UV-optical and optical CMDs are shown in \autoref{fig:cmd}. The BS are generally more massive than the stars currently evolving at the MS turn-off and are located along the brighter and hotter extension of the MS locus, mimicking the location of intermediate-age stars. Therefore, we generated relatively younger ages BaSTI-IAC isochrones by adopting the same assumptions concerning metallicity and alpha-element enhancement as the old stellar component. The comparison between model predictions and the empirical data in the B$-$V vs V and BaF2$-$V vs BaF2 CMDs is shown in \autoref{fig:cmd_bs}. We see that the isochrone with age 3 Gyr is lying closer to the observed BSS. Based on the observed location of the BSS in optical and UV-optical CMDs, we extracted the mass, luminosity, effective temperature of the BSS from the 3 Gyr BaSTI-IAC isochrone which are  $M = 1.14\pm0.04 M_\odot$, $\log L/L_\odot = 0.9\pm0.1$, and $\log T_{\mathrm{eff}}/K = 3.93\pm0.01$, respectively.

\begin{table*}
    \caption{List of the telescopes and their filters used in the SED fit. }
    \label{tab:telescope}
    \centering
    \begin{threeparttable}
    
    \adjustbox{max width=0.75\textwidth} {
    \begin{tabular}{ c c c c }
    \hline
    Telescope & Filters & Wavelength range in \AA &  Reference\\
    \hline
    UVIT/\mbox{{\em AstroSat}} & BaF2, Sapphire & 1350 - 1,800  & This paper \\ 
    {\em GALEX} & FUV, NUV & 1350 - 3000   & \citet{Schiavon2012} \\
    CFHT-3.6m  & U, B, V, R, I & 3000 - 11800  & \citet{Stetson2019} \\
    SDSS  & u, g, r, i, z & 3000 - 10800  & DR9, \citet{Ahn2012} \\
    GAIA  & G, BP & 3300 - 10600  & \citet{GaiaCatalog2018} \\
    PAN-STARRS & g, r, i, z, y & 3900 - 10800  & \citet{Chambers2016} \\
    CTIO/DECam  & g, r &  3925 - 7233  & DECam Legacy Survey DR3 \tnote{a} \\
    \hline
    \end{tabular} }
    
    \begin{tablenotes}
       \item[a] https://www.legacysurvey.org/dr3/description/
    \end{tablenotes}  
 \end{threeparttable} 
\end{table*}

We obtained optical photometry of the FUV bright BSS from various archival catalogs and performed spectral energy distribution (SED) fitting on the observed fluxes to the model generated fluxes for different filters from UV to near-IR bandwidth. We used an online SED fitting tool, VO SED analyszer \citep[VOSA, ][]{Bayo2008}, to perform the SED fit. It uses the chi-square minimisation technique to fit the observed fluxes on the model fluxes generated from the theoretical spectra incorporating various stellar atmosphere models. We used the Kurucz stellar atmosphere model, ATLAS9 \citep[Kurucz model, ][]{Castelli2003} grids (spectra) with model parameters in the following range, T$_{\mathrm{eff}}$: 3,500 K to 50,000 K, $[Fe/H]: -2.0$ and $-1.5$ (nearest to the cluster metallicity), and log (g): 0.0 to 5.0, respectively. The observed fluxes of different filters used in the SED fit is given in \autoref{tab:telescope}. We have used observed fluxes from 23 filters in the wavelength range 1,800 \AA\ to 11,800 \AA\ to perform the SED fit.

\begin{figure*}
    \centering
    \includegraphics[width=0.80\textwidth]{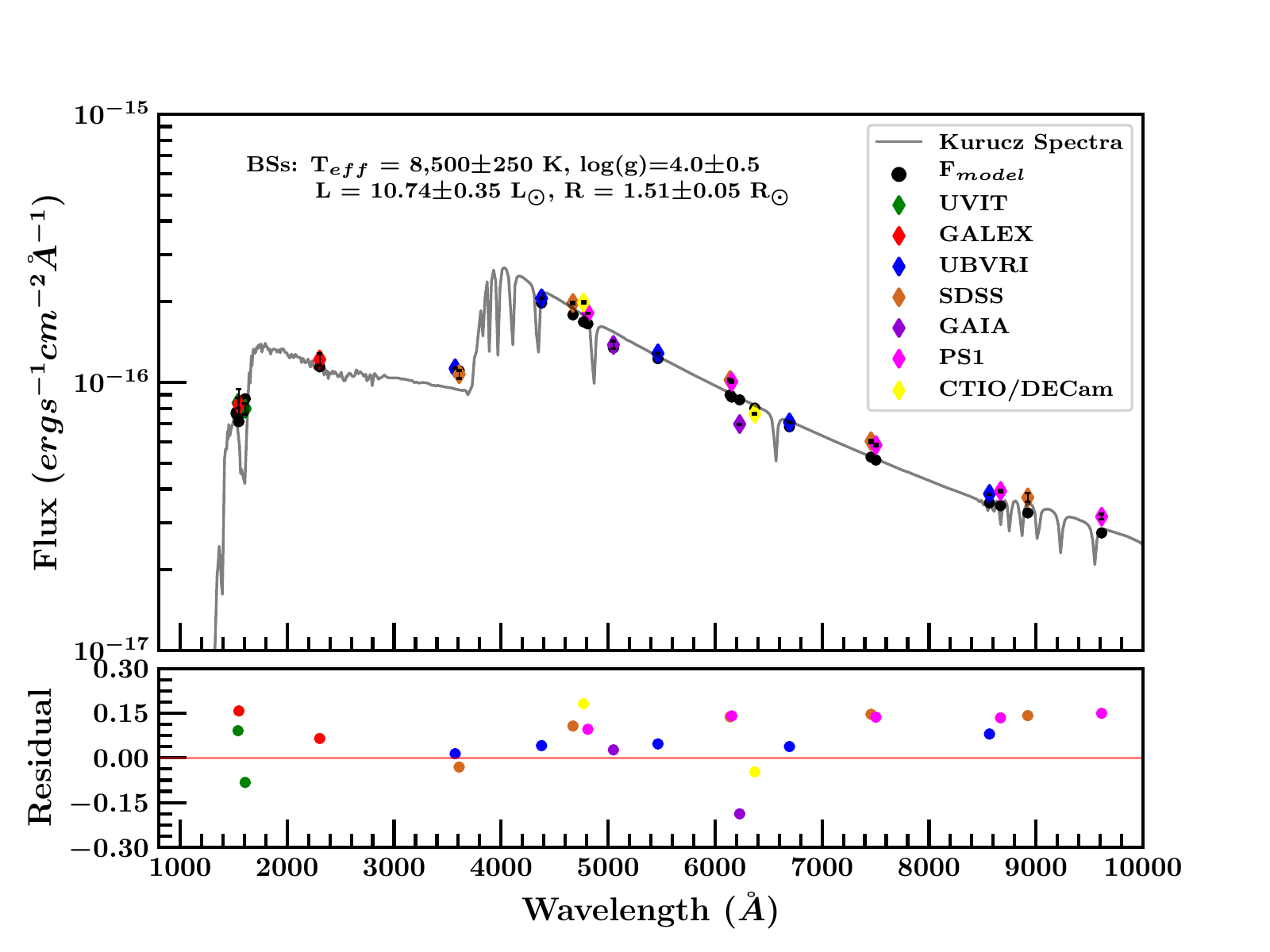}
    \caption{The SED fit of BSS using the Kurucz model is shown in the upper panel. The bottom panel shows the residue (deviation) of the observed fluxes from the Kurucz model fluxes.}
    \label{fig:sed_bs}
\end{figure*}

\begin{table*}
    \caption{Various physical parameters of the BSS derived from the SED fit and BaSTI-IAC isochrones are listed in the table.}
    \label{tab:bss_sed}
    \centering
    
    \adjustbox{max width=\textwidth} {
    \begin{tabular}{ c c c c c c c c c}
    \hline
  ID & RAJ2000 & DEJ2000 & fit & T$_{\mathrm{eff}}$ & Luminosity  & Radius    & log(g) & Mass \\
      & degree  & degree &      & K         & L$_\odot$   & R$_\odot$ &  dex & M$_\odot$ \\
    \hline
 \multirow{ 2}{*}{BSS01} &  \multirow{ 2}{*}{182.5633} & \multirow{ 2}{*}{18.5480} & Isochrone & 8511 $\pm$ 196 & 7.94 $\pm$ 1.83 & 1.30 $\pm$ 0.07 & - & 1.14$\pm$0.04  \\ 
                   &            &              & SED & 8500 $\pm$ 250 & 10.74 $\pm$ 0.35 & 1.51 $\pm$ 0.05 & 4.0$\pm$0.5 & 1.23$\pm$0.02 \\ 
    \hline
    \end{tabular} }
\end{table*}

In \autoref{fig:sed_bs}, we have shown the best fitted Kurucz model spectra of 8,500 K on the observed fluxes in gray line. The model fluxes calculated for various filters using the Kurucz model spectra are shown in solid black circles and the observed fluxes from the UVIT, {\em GALEX}, CFHT-3.6m, SDSS, GAIA, PAN-STARRS, and CTIO/DECam telescopes are shown in green, red, blue, brown, violet, magenta, and yellow diamonds, respectively. In the lower panel  of \autoref{fig:sed_bs}, we have shown the fractional deviation of the observed flux from the model flux and indicated as residual [(= $F_{\mathrm{obs.}} - F_{\mathrm{mod}})/F_{\mathrm{mod}})$]. The best fitted Kurucz model spectra with physical parameters T$_{\mathrm{eff}}$ = 8,500 K, $[Fe/H] = -2.0$ and log (g) = 4.0, was obtained with a reduced chi-square value of 9.3. The deviation of the observed fluxes from the theoretical fluxes for most of the filters is  less than 15\%. But for a few filters (e.g., CTIO/DECam g and Gaia G filters) the deviation is more than or around 15\% which might be the reason for the high value of $\chi^{2}$. However, we can see that the overall photometric magnitudes are well within the maximum photometric errors of 20\% (0.2 mag). We extracted bolometric luminosity and radius of the FUV bright BSS from the slope of the best fitted spectrum using scaling relation $(R/d)^2$ used in the SED fit, where R is the radius of the source and $d$ is the distance. The distance of the cluster = 19.0 kpc and extinction value A$_{\mathrm{v}} = 0.068$ magnitude were used in the SED fit. We have also estimated the mass of the BSS using the T$_{\mathrm{eff}}$, luminosity and radius derived from the SED analysis. The derived physical parameters from the SED fit and the BaSTI-IAC isochrone fitting are listed in \autoref{tab:bss_sed}. We find that the T$_{\mathrm{eff}}$, mass and luminosity of the BSS, derived from the SED fit and BaSTI-IAC isochrone fitting on CMDs are matching well within the error range.  

\section{Summary and Conclusion}
\label{sec:conclusion}
UV analysis of the GGC NGC 4147 is presented using observations in three FUV filters of UVIT. A catalog of the UVIT detected BHBs, and BSS along with their FUV magnitudes in three UVIT filters is provided in \autoref{tab:phot}. It is found that all the FUV bright BHBs belong to the 2G population as defined by \citep{Villanova2016} and they lie in two sub-populations on the UV-optical CMDs with a count ratio of 24:13 (BHB1:BHB2). The derived ranges of T$_{\mathrm{eff}}$ of BHB1 and BHB2 are 10,000 - 11,300 K and 8,000 - 9,500 K, respectively with a gap of 500 K between both the sub-populations. Spectroscopic analysis of the sources is required to investigate further on the distinction between the two sub-population of BHBs. The physical parameters of the FUV bright BSS were derived by fitting younger BaSTI-IAC isochrones in optical and UV-optical CMDs, and SED fitting of the observed fluxes from UV to near-IR wavelengths with the Kurucz model. The derived physical parameters of the FUV bright BSS are listed in \autoref{tab:bss_sed}. 

\vspace{1em}

%%Use section* for acknowledgements
\section*{Acknowledgements}
 We would like to thank the referee for her/his valuable suggestions and comments. RK would like to acknowledge CSIR Research Fellowship (JRF) Grant No.09/983(0034)/2019-EMR-1 for the financial support. ACP would like to acknowledge the support by Indian Space Research Organization, Department of Space, Government of India (ISRO RESPOND project No.ISRO/RES/2/409/17-18). ACP thanks Inter University centre for Astronomy and Astrophysics (IUCAA), Pune, India for providing facilities to carry out his work. DKO acknowledges the support of the Department of Atomic Energy, Government of India, under Project Identification No. RTI 4002. AM thanks DST-INSPIRE (IF150845) for the funding. This publication uses the data from the \mbox{{\em AstroSat}} mission of the Indian Space Research  Organisation (ISRO), archived at the Indian Space Science Data Center (ISSDC). The UVIT data used here was processed by the Payload Operations Centre at IIA. The UVIT is built in collaboration between IIA, IUCAA, TIFR, ISRO and CSA.

%%use \balance somewhere in the left column of the last page to balance the two columns in the end page
\balance

%%References section

\bibliography{revised}{}

\begin{thebibliography}{}
\expandafter\ifx\csname natexlab\endcsname\relax\def\natexlab#1{#1}\fi

\bibitem[{{Ahn} {$et~al$.}(2012){Ahn}, {Alexandroff}, {Allende Prieto},
  {Anderson}, {Anderton}, {Andrews}, {Aubourg}, {Bailey}, {Balbinot}, {Barnes},
  {Bautista}, {Beers}, {Beifiori}, {Berlind}, {Bhardwaj}, {Bizyaev}, {Blake},
  {Blanton}, {Blomqvist}, {Bochanski}, {Bolton}, {Borde}, {Bovy}, {Brandt},
  {Brinkmann}, {Brown}, {Brownstein}, {Bundy}, {Busca}, {Carithers}, {Carnero},
  {Carr}, {Casetti-Dinescu}, {Chen}, {Chiappini}, {Comparat}, {Connolly},
  {Crepp}, {Cristiani}, {Croft}, {Cuesta}, {da Costa}, {Davenport}, {Dawson},
  {de Putter}, {De Lee}, {Delubac}, {Dhital}, {Ealet}, {Ebelke}, {Edmondson},
  {Eisenstein}, {Escoffier}, {Esposito}, {Evans}, {Fan}, {Femen{\'\i}a
  Castell{\'a}}, {Fern{\'a}ndez Alvar}, {Ferreira}, {Filiz Ak}, {Finley},
  {Fleming}, {Font-Ribera}, {Frinchaboy}, {Garc{\'\i}a-Hern{\'a}ndez},
  {Garc{\'\i}a P{\'e}rez}, {Ge}, {G{\'e}nova-Santos}, {Gillespie}, {Girardi},
  {Gonz{\'a}lez Hern{\'a}ndez}, {Grebel}, {Gunn}, {Guo}, {Haggard}, {Hamilton},
  {Harris}, {Hawley}, {Hearty}, {Ho}, {Hogg}, {Holtzman}, {Honscheid},
  {Huehnerhoff}, {Ivans}, {Ivezi{\'c}}, {Jacobson}, {Jiang}, {Johansson},
  {Johnson}, {Kauffmann}, {Kirkby}, {Kirkpatrick}, {Klaene}, {Knapp}, {Kneib},
  {Le Goff}, {Leauthaud}, {Lee}, {Lee}, {Long}, {Loomis}, {Lucatello},
  {Lundgren}, {Lupton}, {Ma}, {Ma}, {MacDonald}, {Mack}, {Mahadevan}, {Maia},
  {Majewski}, {Makler}, {Malanushenko}, {Malanushenko}, {Manchado},
  {Mandelbaum}, {Manera}, {Maraston}, {Margala}, {Martell}, {McBride},
  {McGreer}, {McMahon}, {M{\'e}nard}, {Meszaros}, {Miralda-Escud{\'e}},
  {Montero-Dorta}, {Montesano}, {Morrison}, {Muna}, {Munn}, {Murayama},
  {Myers}, {Neto}, {Nguyen}, {Nichol}, {Nidever}, {Noterdaeme}, {Nuza}, {Ogand
  o}, {Olmstead}, {Oravetz}, {Owen}, {Padmanabhan}, {Palanque-Delabrouille},
  {Pan}, {Parejko}, {Parihar}, {P{\^a}ris}, {Pattarakijwanich}, {Pepper},
  {Percival}, {P{\'e}rez-Fournon}, {P{\'e}rez-R{\`a}fols}, {Petitjean},
  {Pforr}, {Pieri}, {Pinsonneault}, {Porto de Mello}, {Prada}, {Price-Whelan},
  {Raddick}, {Rebolo}, {Rich}, {Richards}, {Robin}, {Rocha-Pinto}, {Rockosi},
  {Roe}, {Ross}, {Ross}, {Rossi}, {Rubi{\~n}o-Martin}, {Samushia}, {Sanchez
  Almeida}, {S{\'a}nchez}, {Santiago}, {Sayres}, {Schlegel}, {Schlesinger},
  {Schmidt}, {Schneider}, {Schultheis}, {Schwope}, {Sc{\'o}ccola}, {Seljak},
  {Sheldon}, {Shen}, {Shu}, {Simmerer}, {Simmons}, {Skibba}, {Skrutskie},
  {Slosar}, {Sobreira}, {Sobeck}, {Stassun}, {Steele}, {Steinmetz}, {Strauss},
  {Streblyanska}, {Suzuki}, {Swanson}, {Tal}, {Thakar}, {Thomas}, {Thompson},
  {Tinker}, {Tojeiro}, {Tremonti}, {Vargas Maga{\~n}a}, {Verde}, {Viel},
  {Vikas}, {Vogt}, {Wake}, {Wang}, {Weaver}, {Weinberg}, {Weiner}, {West},
  {White}, {Wilson}, {Wisniewski}, {Wood-Vasey}, {Yanny}, {Y{\`e}che}, {York},
  {Zamora}, {Zasowski}, {Zehavi}, {Zhao}, {Zheng}, {Zhu}, \& {Zinn}}]{Ahn2012}
{Ahn}, C.~P., {Alexandroff}, R., {Allende Prieto}, C., {$et~al$.} 2012, \apjs,
  203, 21

\bibitem[{{Arellano Ferro} {$et~al$.}(2004){Arellano Ferro}, {Ar{\'e}valo},
  {L{\'a}zaro}, {Rey}, {Bramich}, \& {Giridhar}}]{Arellano2004}
{Arellano Ferro}, A., {Ar{\'e}valo}, M.~J., {L{\'a}zaro}, C., {$et~al$.} 2004,
  \rmxaa, 40, 209

\bibitem[{{Arellano Ferro} {$et~al$.}(2018){Arellano Ferro}, {Rojas Galindo},
  {Muneer}, \& {Giridhar}}]{Arellano2018}
{Arellano Ferro}, A., {Rojas Galindo}, F.~C., {Muneer}, S., \& {Giridhar}, S.
  2018, \rmxaa, 54, 325

\bibitem[{{Auriere} \& {Lauzeral}(1991)}]{Auriere1991}
{Auriere}, M., \& {Lauzeral}, C. 1991, \aap, 244, 303

\bibitem[{{Bastian} \& {Lardo}(2018)}]{Bastian2018}
{Bastian}, N., \& {Lardo}, C. 2018, \araa, 56, 83

\bibitem[{{Bayo} {$et~al$.}(2008){Bayo}, {Rodrigo}, {Barrado Y Navascu{\'e}s},
  {Solano}, {Guti{\'e}rrez}, {Morales-Calder{\'o}n}, \& {Allard}}]{Bayo2008}
{Bayo}, A., {Rodrigo}, C., {Barrado Y Navascu{\'e}s}, D., {$et~al$.} 2008,
  \aap, 492, 277

\bibitem[{{Behr} {$et~al$.}(2000){Behr}, {Djorgovski}, {Cohen}, {McCarthy},
  {C{\^o}t{\'e}}, {Piotto}, \& {Zoccali}}]{Behr2000}
{Behr}, B.~B., {Djorgovski}, S.~G., {Cohen}, J.~G., {$et~al$.} 2000, \apj, 528,
  849

\bibitem[{{Brown} {$et~al$.}(2016){Brown}, {Cassisi}, {D'Antona}, {Salaris},
  {Milone}, {Dalessandro}, {Piotto}, {Renzini}, {Sweigart}, {Bellini},
  {Ortolani}, {Sarajedini}, {Aparicio}, {Bedin}, {Anderson}, {Pietrinferni}, \&
  {Nardiello}}]{Brown2016}
{Brown}, T.~M., {Cassisi}, S., {D'Antona}, F., {$et~al$.} 2016, \apj, 822, 44

\bibitem[{{Cardelli} {$et~al$.}(1989){Cardelli}, {Clayton}, \&
  {Mathis}}]{cardeli1989}
{Cardelli}, J.~A., {Clayton}, G.~C., \& {Mathis}, J.~S. 1989, \apj, 345, 245

\bibitem[{{Cassisi} \& {Salaris}(2020)}]{Cassisi2020}
{Cassisi}, S., \& {Salaris}, M. 2020, \aapr, 28, 5

\bibitem[{{Castelli} \& {Kurucz}(2003)}]{Castelli2003}
{Castelli}, F., \& {Kurucz}, R.~L. 2003, in Modelling of Stellar Atmospheres,
  ed. N.~{Piskunov}, W.~W. {Weiss}, \& D.~F. {Gray}, Vol. 210, A20

\bibitem[{{Catelan} {$et~al$.}(1998){Catelan}, {Borissova}, {Sweigart}, \&
  {Spassova}}]{Catelan1998}
{Catelan}, M., {Borissova}, J., {Sweigart}, A.~V., \& {Spassova}, N. 1998,
  \apj, 494, 265

\bibitem[{{Chambers} {$et~al$.}(2016){Chambers}, {Magnier}, {Metcalfe},
  {Flewelling}, {Huber}, {Waters}, {Denneau}, {Draper}, {Farrow}, {Finkbeiner},
  {Holmberg}, {Koppenhoefer}, {Price}, {Rest}, {Saglia}, {Schlafly}, {Smartt},
  {Sweeney}, {Wainscoat}, {Burgett}, {Chastel}, {Grav}, {Heasley}, {Hodapp},
  {Jedicke}, {Kaiser}, {Kudritzki}, {Luppino}, {Lupton}, {Monet}, {Morgan},
  {Onaka}, {Shiao}, {Stubbs}, {Tonry}, {White}, {Ba{\~n}ados}, {Bell},
  {Bender}, {Bernard}, {Boegner}, {Boffi}, {Botticella}, {Calamida},
  {Casertano}, {Chen}, {Chen}, {Cole}, {Deacon}, {Frenk}, {Fitzsimmons},
  {Gezari}, {Gibbs}, {Goessl}, {Goggia}, {Gourgue}, {Goldman}, {Grant},
  {Grebel}, {Hambly}, {Hasinger}, {Heavens}, {Heckman}, {Henderson}, {Henning},
  {Holman}, {Hopp}, {Ip}, {Isani}, {Jackson}, {Keyes}, {Koekemoer}, {Kotak},
  {Le}, {Liska}, {Long}, {Lucey}, {Liu}, {Martin}, {Masci}, {McLean}, {Mindel},
  {Misra}, {Morganson}, {Murphy}, {Obaika}, {Narayan}, {Nieto-Santisteban},
  {Norberg}, {Peacock}, {Pier}, {Postman}, {Primak}, {Rae}, {Rai}, {Riess},
  {Riffeser}, {Rix}, {R{\"o}ser}, {Russel}, {Rutz}, {Schilbach}, {Schultz},
  {Scolnic}, {Strolger}, {Szalay}, {Seitz}, {Small}, {Smith}, {Soderblom},
  {Taylor}, {Thomson}, {Taylor}, {Thakar}, {Thiel}, {Thilker}, {Unger},
  {Urata}, {Valenti}, {Wagner}, {Walder}, {Walter}, {Watters}, {Werner},
  {Wood-Vasey}, \& {Wyse}}]{Chambers2016}
{Chambers}, K.~C., {Magnier}, E.~A., {Metcalfe}, N., {$et~al$.} 2016, arXiv
  e-prints, arXiv:1612.05560

\bibitem[{{Chatterjee} {$et~al$.}(2013){Chatterjee}, {Rasio}, {Sills}, \&
  {Glebbeek}}]{Chatterjee2013}
{Chatterjee}, S., {Rasio}, F.~A., {Sills}, A., \& {Glebbeek}, E. 2013, \apj,
  777, 106

\bibitem[{{Clement}(2017)}]{Clement2017}
{Clement}, C. 2017, in European Physical Journal Web of Conferences, Vol. 152,
  European Physical Journal Web of Conferences, 01021

\bibitem[{{Dieball} {$et~al$.}(2010){Dieball}, {Long}, {Knigge}, {Thomson}, \&
  {Zurek}}]{Dieball2010}
{Dieball}, A., {Long}, K.~S., {Knigge}, C., {Thomson}, G.~S., \& {Zurek}, D.~R.
  2010, \apj, 710, 332

\bibitem[{{Ferraro} {$et~al$.}(2003){Ferraro}, {Sills}, {Rood}, {Paltrinieri},
  \& {Buonanno}}]{Ferraro2003}
{Ferraro}, F.~R., {Sills}, A., {Rood}, R.~T., {Paltrinieri}, B., \& {Buonanno},
  R. 2003, \apj, 588, 464

\bibitem[{{Ferraro} {$et~al$.}(1997){Ferraro}, {Paltrinieri}, {Fusi Pecci},
  {Cacciari}, {Dorman}, {Rood}, {Buonanno}, {Corsi}, {Burgarella}, \&
  {Laget}}]{Ferraro1997}
{Ferraro}, F.~R., {Paltrinieri}, B., {Fusi Pecci}, F., {$et~al$.} 1997, \aap,
  324, 915

\bibitem[{{Gaia Collaboration} {$et~al$.}(2018){Gaia Collaboration}, {Brown},
  {Vallenari}, {Prusti}, {de Bruijne}, {Babusiaux}, {Bailer-Jones}, {Biermann},
  {Evans}, {Eyer}, {Jansen}, {Jordi}, {Klioner}, {Lammers}, {Lindegren},
  {Luri}, {Mignard}, {Panem}, {Pourbaix}, {Randich}, {Sartoretti}, {Siddiqui},
  {Soubiran}, {van Leeuwen}, {Walton}, {Arenou}, {Bastian}, {Cropper},
  {Drimmel}, {Katz}, {Lattanzi}, {Bakker}, {Cacciari}, {Casta{\~n}eda},
  {Chaoul}, {Cheek}, {De Angeli}, {Fabricius}, {Guerra}, {Holl}, {Masana},
  {Messineo}, {Mowlavi}, {Nienartowicz}, {Panuzzo}, {Portell}, {Riello},
  {Seabroke}, {Tanga}, {Th{\'e}venin}, {Gracia-Abril}, {Comoretto},
  {Garcia-Reinaldos}, {Teyssier}, {Altmann}, {Andrae}, {Audard},
  {Bellas-Velidis}, {Benson}, {Berthier}, {Blomme}, {Burgess}, {Busso},
  {Carry}, {Cellino}, {Clementini}, {Clotet}, {Creevey}, {Davidson}, {De
  Ridder}, {Delchambre}, {Dell'Oro}, {Ducourant},
  {Fern{\'a}ndez-Hern{\'a}ndez}, {Fouesneau}, {Fr{\'e}mat}, {Galluccio},
  {Garc{\'\i}a-Torres}, {Gonz{\'a}lez-N{\'u}{\~n}ez}, {Gonz{\'a}lez-Vidal},
  {Gosset}, {Guy}, {Halbwachs}, {Hambly}, {Harrison}, {Hern{\'a}ndez},
  {Hestroffer}, {Hodgkin}, {Hutton}, {Jasniewicz}, {Jean-Antoine-Piccolo},
  {Jordan}, {Korn}, {Krone-Martins}, {Lanzafame}, {Lebzelter}, {L{\"o}ffler},
  {Manteiga}, {Marrese}, {Mart{\'\i}n-Fleitas}, {Moitinho}, {Mora}, {Muinonen},
  {Osinde}, {Pancino}, {Pauwels}, {Petit}, {Recio-Blanco}, {Richards},
  {Rimoldini}, {Robin}, {Sarro}, {Siopis}, {Smith}, {Sozzetti}, {S{\"u}veges},
  {Torra}, {van Reeven}, {Abbas}, {Abreu Aramburu}, {Accart}, {Aerts},
  {Altavilla}, {{\'A}lvarez}, {Alvarez}, {Alves}, {Anderson}, {Andrei},
  {Anglada Varela}, {Antiche}, {Antoja}, {Arcay}, {Astraatmadja}, {Bach},
  {Baker}, {Balaguer-N{\'u}{\~n}ez}, {Balm}, {Barache}, {Barata}, {Barbato},
  {Barblan}, {Barklem}, {Barrado}, {Barros}, {Barstow}, {Bartholom{\'e}
  Mu{\~n}oz}, {Bassilana}, {Becciani}, {Bellazzini}, {Berihuete}, {Bertone},
  {Bianchi}, {Bienaym{\'e}}, {Blanco-Cuaresma}, {Boch}, {Boeche}, {Bombrun},
  {Borrachero}, {Bossini}, {Bouquillon}, {Bourda}, {Bragaglia}, {Bramante},
  {Breddels}, {Bressan}, {Brouillet}, {Br{\"u}semeister}, {Brugaletta},
  {Bucciarelli}, {Burlacu}, {Busonero}, {Butkevich}, {Buzzi}, {Caffau},
  {Cancelliere}, {Cannizzaro}, {Cantat-Gaudin}, {Carballo}, {Carlucci},
  {Carrasco}, {Casamiquela}, {Castellani}, {Castro-Ginard}, {Charlot},
  {Chemin}, {Chiavassa}, {Cocozza}, {Costigan}, {Cowell}, {Crifo}, {Crosta},
  {Crowley}, {Cuypers}, {Dafonte}, {Damerdji}, {Dapergolas}, {David}, {David},
  {de Laverny}, {De Luise}, {De March}, {de Martino}, {de Souza}, {de Torres},
  {Debosscher}, {del Pozo}, {Delbo}, {Delgado}, {Delgado}, {Di Matteo},
  {Diakite}, {Diener}, {Distefano}, {Dolding}, {Drazinos}, {Dur{\'a}n},
  {Edvardsson}, {Enke}, {Eriksson}, {Esquej}, {Eynard Bontemps}, {Fabre},
  {Fabrizio}, {Faigler}, {Falc{\~a}o}, {Farr{\`a}s Casas}, {Federici},
  {Fedorets}, {Fernique}, {Figueras}, {Filippi}, {Findeisen}, {Fonti},
  {Fraile}, {Fraser}, {Fr{\'e}zouls}, {Gai}, {Galleti}, {Garabato},
  {Garc{\'\i}a-Sedano}, {Garofalo}, {Garralda}, {Gavel}, {Gavras}, {Gerssen},
  {Geyer}, {Giacobbe}, {Gilmore}, {Girona}, {Giuffrida}, {Glass}, {Gomes},
  {Granvik}, {Gueguen}, {Guerrier}, {Guiraud}, {Guti{\'e}rrez-S{\'a}nchez},
  {Haigron}, {Hatzidimitriou}, {Hauser}, {Haywood}, {Heiter}, {Helmi}, {Heu},
  {Hilger}, {Hobbs}, {Hofmann}, {Holland}, {Huckle}, {Hypki}, {Icardi},
  {Jan{\ss}en}, {Jevardat de Fombelle}, {Jonker}, {Juh{\'a}sz}, {Julbe},
  {Karampelas}, {Kewley}, {Klar}, {Kochoska}, {Kohley}, {Kolenberg},
  {Kontizas}, {Kontizas}, {Koposov}, {Kordopatis}, {Kostrzewa-Rutkowska},
  {Koubsky}, {Lambert}, {Lanza}, {Lasne}, {Lavigne}, {Le Fustec}, {Le
  Poncin-Lafitte}, {Lebreton}, {Leccia}, {Leclerc}, {Lecoeur-Taibi},
  {Lenhardt}, {Leroux}, {Liao}, {Licata}, {Lindstr{\o}m}, {Lister}, {Livanou},
  {Lobel}, {L{\'o}pez}, {Managau}, {Mann}, {Mantelet}, {Marchal}, {Marchant},
  {Marconi}, {Marinoni}, {Marschalk{\'o}}, {Marshall}, {Martino}, {Marton},
  {Mary}, {Massari}, {Matijevi{\v{c}}}, {Mazeh}, {McMillan}, {Messina},
  {Michalik}, {Millar}, {Molina}, {Molinaro}, {Moln{\'a}r}, {Montegriffo},
  {Mor}, {Morbidelli}, {Morel}, {Morris}, {Mulone}, {Muraveva}, {Musella},
  {Nelemans}, {Nicastro}, {Noval}, {O'Mullane}, {Ord{\'e}novic},
  {Ord{\'o}{\~n}ez-Blanco}, {Osborne}, {Pagani}, {Pagano}, {Pailler},
  {Palacin}, {Palaversa}, {Panahi}, {Pawlak}, {Piersimoni}, {Pineau}, {Plachy},
  {Plum}, {Poggio}, {Poujoulet}, {Pr{\v{s}}a}, {Pulone}, {Racero}, {Ragaini},
  {Rambaux}, {Ramos-Lerate}, {Regibo}, {Reyl{\'e}}, {Riclet}, {Ripepi}, {Riva},
  {Rivard}, {Rixon}, {Roegiers}, {Roelens}, {Romero-G{\'o}mez}, {Rowell},
  {Royer}, {Ruiz-Dern}, {Sadowski}, {Sagrist{\`a} Sell{\'e}s}, {Sahlmann},
  {Salgado}, {Salguero}, {Sanna}, {Santana-Ros}, {Sarasso}, {Savietto},
  {Schultheis}, {Sciacca}, {Segol}, {Segovia}, {S{\'e}gransan}, {Shih},
  {Siltala}, {Silva}, {Smart}, {Smith}, {Solano}, {Solitro}, {Sordo}, {Soria
  Nieto}, {Souchay}, {Spagna}, {Spoto}, {Stampa}, {Steele},
  {Steidelm{\"u}ller}, {Stephenson}, {Stoev}, {Suess}, {Surdej}, {Szabados},
  {Szegedi-Elek}, {Tapiador}, {Taris}, {Tauran}, {Taylor}, {Teixeira},
  {Terrett}, {Teyssand ier}, {Thuillot}, {Titarenko}, {Torra Clotet}, {Turon},
  {Ulla}, {Utrilla}, {Uzzi}, {Vaillant}, {Valentini}, {Valette}, {van Elteren},
  {Van Hemelryck}, {van Leeuwen}, {Vaschetto}, {Vecchiato}, {Veljanoski},
  {Viala}, {Vicente}, {Vogt}, {von Essen}, {Voss}, {Votruba}, {Voutsinas},
  {Walmsley}, {Weiler}, {Wertz}, {Wevers}, {Wyrzykowski}, {Yoldas},
  {{\v{Z}}erjal}, {Ziaeepour}, {Zorec}, {Zschocke}, {Zucker}, {Zurbach}, \&
  {Zwitter}}]{GaiaCatalog2018}
{Gaia Collaboration}, {Brown}, A.~G.~A., {Vallenari}, A., {$et~al$.} 2018,
  \aap, 616, A1

\bibitem[{{Gosnell} {$et~al$.}(2015){Gosnell}, {Mathieu}, {Geller}, {Sills},
  {Leigh}, \& {Knigge}}]{Gosnell2015}
{Gosnell}, N.~M., {Mathieu}, R.~D., {Geller}, A.~M., {$et~al$.} 2015, \apj,
  814, 163

\bibitem[{{Gratton} {$et~al$.}(2019){Gratton}, {Bragaglia}, {Carretta},
  {D'Orazi}, {Lucatello}, \& {Sollima}}]{Gratton2019}
{Gratton}, R., {Bragaglia}, A., {Carretta}, E., {$et~al$.} 2019, \aapr, 27, 8

\bibitem[{{Harris}(2010)}]{Harris2010}
{Harris}, W.~E. 2010, arXiv e-prints, arXiv:1012.3224

\bibitem[{{Hidalgo} {$et~al$.}(2018){Hidalgo}, {Pietrinferni}, {Cassisi},
  {Salaris}, {Mucciarelli}, {Savino}, {Aparicio}, {Silva Aguirre}, \&
  {Verma}}]{Hidalgo2018}
{Hidalgo}, S.~L., {Pietrinferni}, A., {Cassisi}, S., {$et~al$.} 2018, \apj,
  856, 125

\bibitem[{{Ivans}(2009)}]{Ivans2009}
{Ivans}, I.~I. 2009, in American Astronomical Society Meeting Abstracts, Vol.
  213, American Astronomical Society Meeting Abstracts \#213, 341.06

\bibitem[{{Jadhav} {$et~al$.}(2019){Jadhav}, {Sindhu}, \&
  {Subramaniam}}]{Jadhav2019}
{Jadhav}, V.~V., {Sindhu}, N., \& {Subramaniam}, A. 2019, \apj, 886, 13

\bibitem[{{Jain} {$et~al$.}(2019){Jain}, {Vig}, \& {Ghosh}}]{Jain2019}
{Jain}, R., {Vig}, S., \& {Ghosh}, S.~K. 2019, \mnras, 485, 2877

\bibitem[{{Knigge} {$et~al$.}(2009){Knigge}, {Leigh}, \& {Sills}}]{Knigge2009}
{Knigge}, C., {Leigh}, N., \& {Sills}, A. 2009, \nat, 457, 288

\bibitem[{{Kumar} {$et~al$.}(2012){Kumar}, {Ghosh}, \& et~al.}]{Kumar2012}
{Kumar}, A., {Ghosh}, S.~K., \& et~al. 2012, in \procspie, Vol. 8443, Space
  Telescopes and Instrumentation 2012: Ultraviolet to Gamma Ray, 84431N

\bibitem[{{Kumar} {$et~al$.}(2020{\natexlab{a}}){Kumar}, {Pradhan},
  {Mohapatra}, {Moharana}, {Parthasarathy}, {Ojha}, \& {Murthy}}]{Kumar2020a}
{Kumar}, R., {Pradhan}, A.~C., {Mohapatra}, A., {$et~al$.} 2020{\natexlab{a}},
  \mnras, revision submitted

\bibitem[{{Kumar} {$et~al$.}(2020{\natexlab{b}}){Kumar}, {Pradhan},
  {Parthasarathy}, {Ojha}, {Mohapatra}, \& {Murthy}}]{Kumar2020}
{Kumar}, R., {Pradhan}, A.~C., {Parthasarathy}, M., {$et~al$.}
  2020{\natexlab{b}}, in Star Clusters: From the Milky Way to the Early
  Universe, ed. A.~{Bragaglia}, M.~{Davies}, A.~{Sills}, \& E.~{Vesperini},
  Vol. 351, 464--467

\bibitem[{{Lata} {$et~al$.}(2019){Lata}, {Pandey}, {Pandey}, {Yadav}, {Pandey},
  {Gupta}, {Bangia}, {Chand}, {Jaiswar}, {Joshi}, {Joshi}, {Kumar}, {Kumar},
  {Medhi}, {Misra}, {Nanjappa}, {Pant}, {Purushottam}, {Krishna Reddy}, {Sahu},
  {Sharma}, {Uddin}, \& {Yadav}}]{Lata2019}
{Lata}, S., {Pandey}, A.~K., {Pandey}, J.~C., {$et~al$.} 2019, \aj, 158, 51

\bibitem[{{Leigh} {$et~al$.}(2013){Leigh}, {Knigge}, {Sills}, {Perets},
  {Sarajedini}, \& {Glebbeek}}]{Leigh2013}
{Leigh}, N., {Knigge}, C., {Sills}, A., {$et~al$.} 2013, \mnras, 428, 897

\bibitem[{{Marino} {$et~al$.}(2019){Marino}, {Milone}, {Renzini}, {D'Antona},
  {Anderson}, {Bedin}, {Bellini}, {Cordoni}, {Lagioia}, {Piotto}, \&
  {Tailo}}]{Marino2019}
{Marino}, A.~F., {Milone}, A.~P., {Renzini}, A., {$et~al$.} 2019, \mnras, 487,
  3815

\bibitem[{{Martell} {$et~al$.}(2008){Martell}, {Smith}, \&
  {Briley}}]{Martell2008}
{Martell}, S.~L., {Smith}, G.~H., \& {Briley}, M.~M. 2008, \aj, 136, 2522

\bibitem[{{Milone} {$et~al$.}(2017){Milone}, {Piotto}, {Renzini}, {Marino},
  {Bedin}, {Vesperini}, {D'Antona}, {Nardiello}, {Anderson}, {King}, {Yong},
  {Bellini}, {Aparicio}, {Barbuy}, {Brown}, {Cassisi}, {Ortolani}, {Salaris},
  {Sarajedini}, \& {van der Marel}}]{Milone2017}
{Milone}, A.~P., {Piotto}, G., {Renzini}, A., {$et~al$.} 2017, \mnras, 464,
  3636

\bibitem[{{Nardiello} {$et~al$.}(2018){Nardiello}, {Libralato}, {Piotto},
  {Anderson}, {Bellini}, {Aparicio}, {Bedin}, {Cassisi}, {Granata}, {King},
  {Lucertini}, {Marino}, {Milone}, {Ortolani}, {Platais}, \& {van der
  Marel}}]{Nardiello2018}
{Nardiello}, D., {Libralato}, M., {Piotto}, G., {$et~al$.} 2018, \mnras, 481,
  3382

\bibitem[{{Pietrinferni} {$et~al$.}(2020){Pietrinferni}, {Hidalgo}, {Cassisi},
  Salaris, Savino, {Mucciarelli}, Verma, Silva~Aguirre, Aparicio, \&
  Ferguson}]{Pietrinferni2020}
{Pietrinferni}, A., {Hidalgo}, S., {Cassisi}, S., {$et~al$.} 2020, \apj,
  submitted

\bibitem[{{Piotto} {$et~al$.}(1999){Piotto}, {Zoccali}, {King}, {Djorgovski},
  {Sosin}, {Rich}, \& {Meylan}}]{Piotto1999}
{Piotto}, G., {Zoccali}, M., {King}, I.~R., {$et~al$.} 1999, \aj, 118, 1727

\bibitem[{Postma \& Leahy(2017)}]{Postma2017}
Postma, J.~E., \& Leahy, D. 2017, \pasp, 129, 115002

\bibitem[{{Rani} {$et~al$.}(2020){Rani}, {Pandey}, {Subramaniam}, {Sahu}, \&
  {Kameswara Rao}}]{Rani2020}
{Rani}, S., {Pandey}, G., {Subramaniam}, A., {Sahu}, S., \& {Kameswara Rao}, N.
  2020, arXiv e-prints, arXiv:2010.00569

\bibitem[{{Sahu} {$et~al$.}(2019{\natexlab{a}}){Sahu}, {Subramaniam},
  {C{\^o}t{\'e}}, {Rao}, \& {Stetson}}]{Sahu2019a}
{Sahu}, S., {Subramaniam}, A., {C{\^o}t{\'e}}, P., {Rao}, N.~K., \& {Stetson},
  P.~B. 2019{\natexlab{a}}, \mnras, 482, 1080

\bibitem[{{Sahu} {$et~al$.}(2019{\natexlab{b}}){Sahu}, {Subramaniam},
  {Simunovic}, {Postma}, {C{\^o}t{\'e}}, {Kameswera Rao}, {Geller}, {Leigh},
  {Shara}, {Puzia}, \& {Stetson}}]{Sahu2019b}
{Sahu}, S., {Subramaniam}, A., {Simunovic}, M., {$et~al$.} 2019{\natexlab{b}},
  \apj, 876, 34

\bibitem[{{Schiavon} {$et~al$.}(2012){Schiavon}, {Dalessandro}, {Sohn}, {Rood},
  {O'Connell}, {Ferraro}, {Lanzoni}, {Beccari}, {Rey}, {Rhee}, {Rich}, {Yoon},
  \& {Lee}}]{Schiavon2012}
{Schiavon}, R.~P., {Dalessandro}, E., {Sohn}, S.~T., {$et~al$.} 2012, \aj, 143,
  121

\bibitem[{{Schlafly} \& {Finkbeiner}(2011)}]{Schlafly2011}
{Schlafly}, E.~F., \& {Finkbeiner}, D.~P. 2011, \apj, 737, 103

\bibitem[{{Sindhu} {$et~al$.}(2019){Sindhu}, {Subramaniam}, {Jadhav},
  {Chatterjee}, {Geller}, {Knigge}, {Leigh}, {Puzia}, {Shara}, \&
  {Simunovic}}]{Sindhu2019}
{Sindhu}, N., {Subramaniam}, A., {Jadhav}, V.~V., {$et~al$.} 2019, \apj, 882,
  43

\bibitem[{{Singh} {$et~al$.}(2020){Singh}, {Sahu}, {Subramaniam}, \&
  {Yadav}}]{Singh2020}
{Singh}, G., {Sahu}, S., {Subramaniam}, A., \& {Yadav}, R.~K.~S. 2020, arXiv
  e-prints, arXiv:2010.06979

\bibitem[{Stetson(1987)}]{Stetson1987}
Stetson, P.~B. 1987, \pasp, 99, 191

\bibitem[{{Stetson} {$et~al$.}(2005){Stetson}, {Catelan}, \&
  {Smith}}]{Stetson2005}
{Stetson}, P.~B., {Catelan}, M., \& {Smith}, H.~A. 2005, \pasp, 117, 1325

\bibitem[{{Stetson} {$et~al$.}(2019){Stetson}, {Pancino}, {Zocchi}, {Sanna}, \&
  {Monelli}}]{Stetson2019}
{Stetson}, P.~B., {Pancino}, E., {Zocchi}, A., {Sanna}, N., \& {Monelli}, M.
  2019, \mnras, 485, 3042

\bibitem[{{Subramaniam} {$et~al$.}(2016){Subramaniam}, {Sindhu}, {Tandon},
  {Kameswara Rao}, {Postma}, {C{\^o}t{\'e}}, {Hutchings}, {Ghosh}, {George},
  {Girish}, {Mohan}, {Murthy}, {Sankarasubramanian}, {Stalin}, {Sutaria},
  {Mondal}, \& {Sahu}}]{Subramaniam2016}
{Subramaniam}, A., {Sindhu}, N., {Tandon}, S.~N., {$et~al$.} 2016, \apjl, 833,
  L27

\bibitem[{{Subramaniam} {$et~al$.}(2017){Subramaniam}, {Sahu}, {Postma},
  {C{\^o}t{\'e}}, {Hutchings}, {Darukhanawalla}, {Chung}, {Tandon}, {Rao},
  {George}, {Ghosh}, {Girish}, {Mohan}, {Murthy}, {Pati}, {Sankarasubramanian},
  {Stalin}, \& {Choudhury}}]{Subramaniam2017}
{Subramaniam}, A., {Sahu}, S., {Postma}, J.~E., {$et~al$.} 2017, \aj, 154, 233

\bibitem[{{Suntzeff} {$et~al$.}(1988){Suntzeff}, {Kraft}, \&
  {Kinman}}]{Suntzeff1988}
{Suntzeff}, N.~B., {Kraft}, R.~P., \& {Kinman}, T.~D. 1988, \aj, 95, 91

\bibitem[{{Tailo} {$et~al$.}(2020){Tailo}, {Milone}, {Lagioia}, {D'Antona},
  {Marino}, {Vesperini}, {Caloi}, {Ventura}, {Dondoglio}, \&
  {Cordoni}}]{Tailo2020}
{Tailo}, M., {Milone}, A.~P., {Lagioia}, E.~P., {$et~al$.} 2020, \mnras, 498,
  5745

\bibitem[{{Tandon} {$et~al$.}(2017){Tandon}, {Subramaniam}, {Girish}, \& et.
  al.}]{Tandon2017}
{Tandon}, S.~N., {Subramaniam}, A., {Girish}, V., \& et. al. 2017, \aj, 154,
  128

\bibitem[{{Tandon} {$et~al$.}(2020){Tandon}, {Postma}, {Joseph}, {Devaraj},
  {Subramaniam}, {Barve}, {George}, {Ghosh}, {Girish}, {Hutchings}, {Kamath},
  {Kathiravan}, {Kumar}, {Lancelot}, {Leahy}, {Mahesh}, {Mohan},
  {Nagabhushana}, {Pati}, {Rao}, {Sankarasubramanian}, {Sriram}, \&
  {Stalin}}]{Tandon2020}
{Tandon}, S.~N., {Postma}, J., {Joseph}, P., {$et~al$.} 2020, \aj, 159, 158

\bibitem[{{Villanova} {$et~al$.}(2016){Villanova}, {Monaco}, {Moni Bidin}, \&
  {Assmann}}]{Villanova2016}
{Villanova}, S., {Monaco}, L., {Moni Bidin}, C., \& {Assmann}, P. 2016, \mnras,
  460, 2351

\end{thebibliography}

\begin{table*}
\caption{UVIT photometry table of the BHB1, BHB2 and FUV bright BSS containing their respective positions, the extinction-corrected magnitudes and the magnitude errors. IDs HB01 to HB24 belong to the BHB1 group and IDs HB25 to HB37 belong to the BHB2 group. The BSS is listed at the end of the table with ID BSS01. }
\label{tab:phot}
\resizebox{\textwidth}{!}{
\begin{tabular}{c c c c c c c c c}
\hline
  \multicolumn{1}{c}{ID} &
  \multicolumn{1}{c}{RAJ2000} &
  \multicolumn{1}{c}{DEJ2000} &
  \multicolumn{1}{c}{BaF2} &
  \multicolumn{1}{c}{eBaF2} &
  \multicolumn{1}{c}{Sapphire} &
  \multicolumn{1}{c}{eSapphire} &
  \multicolumn{1}{c}{Silica} &
  \multicolumn{1}{c}{eSilica} \\
\hline
HB01 &  182.5358 & 18.5215 & 19.57 & 0.01 & 19.37 & 0.02 & 19.35 & 0.02\\
HB02 &  182.5117 & 18.5330 & 19.79 & 0.02 & 19.62 & 0.02 & 19.44 & 0.02\\
HB03 &  182.5076 & 18.5353 & 19.83 & 0.02 & 19.55 & 0.02 & 19.41 & 0.02\\
HB04 &  182.5239 & 18.5356 & 19.43 & 0.01 & 19.23 & 0.02 & 19.55 & 0.03\\
HB05 &  182.5231 & 18.5393 & 19.46 & 0.01 & 19.13 & 0.01 & 19.31 & 0.03\\
HB06 &  182.5243 & 18.5396 & 19.62 & 0.01 & 19.48 & 0.02 & 19.29 & 0.03\\
HB07 &  182.5302 & 18.5413 & 19.45 & 0.01 & 19.27 & 0.01 & 19.45 & 0.02\\
HB08 &  182.5229 & 18.5422 & 19.53 & 0.01 & 19.22 & 0.01 & 19.23 & 0.03\\
HB09 &  182.5309 & 18.5458 & 20.05 & 0.01 & 19.84 & 0.02 & 20.08 & 0.03\\
HB10 &  182.5159 & 18.5455 & 19.41 & 0.02 & 19.13 & 0.02 & 19.15 & 0.02\\
HB11 &  182.5288 & 18.5465 & 19.70 & 0.01 & 19.47 & 0.02 & 19.31 & 0.03\\
HB12 &  182.5297 & 18.5473 & 19.97 & 0.01 & 19.59 & 0.02 & 19.70 & 0.03\\
HB13 &  182.5111 & 18.5564 & 19.95 & 0.01 & 20.07 & 0.02 & 19.73 & 0.02\\
HB14 &  182.5240 & 18.5567 & 19.85 & 0.02 & 19.40 & 0.01 & 19.84 & 0.03\\
HB15 &  182.5394 & 18.5611 & 19.74 & 0.01 & 19.65 & 0.01 & 19.41 & 0.03\\
HB16 &  182.5644 & 18.5639 & 19.49 & 0.01 & 19.42 & 0.02 & 19.44 & 0.02\\
HB17 &  182.5427 & 18.6696 & 19.52 & 0.01 & 19.23 & 0.02 & 19.16 & 0.02\\
HB18 &  182.5103 & 18.5344 & 19.54 & 0.02 & 19.26 & 0.01 & 19.33 & 0.03\\
HB19 &  182.5291 & 18.5376 & 19.22 & 0.03 & 19.09 & 0.03 & 18.98 & 0.03\\
HB20 &  182.5228 & 18.5379 & 19.51 & 0.01 & 19.22 & 0.01 & 19.09 & 0.03\\
HB21 &  182.5315 & 18.5503 & 19.49 & 0.01 & 19.43 & 0.02 & 19.36 & 0.02\\
HB22 &  182.5528 & 18.5609 & 19.36 & 0.01 & 19.11 & 0.01 & 19.33 & 0.02\\
HB23 &  182.5381 & 18.5407 & 19.28 & 0.01 & 19.14 & 0.01 & 19.21 & 0.03\\
HB24 &  182.5279 & 18.5506 & 19.44 & 0.02 & 19.23 & 0.04 & 19.38 & 0.02\\
HB25 &  182.5417 & 18.5302 & 20.63 & 0.03 & 20.54 & 0.03 & 20.05 & 0.03\\
HB26 &  182.5150 & 18.5352 & 21.78 & 0.04 & 21.13 & 0.03 & 20.53 & 0.03\\
HB27 &  182.5268 & 18.5388 & 20.77 & 0.02 & 20.60 & 0.03 & 20.26 & 0.04\\
HB28 &  182.5312 & 18.5420 & 21.26 & 0.03 & 20.84 & 0.03 & 20.82 & 0.06\\
HB29 &  182.5171 & 18.5441 & 21.84 & 0.04 & 21.21 & 0.04 & 20.26 & 0.04\\
HB30 &  182.5097 & 18.5476 & 20.59 & 0.02 & 20.15 & 0.02 & 19.89 & 0.05\\
HB31 &  182.5226 & 18.5489 & 20.75 & 0.03 & 20.13 & 0.02 & 19.91 & 0.02\\
HB32 &  182.4897 & 18.5769 & 21.73 & 0.04 & 21.17 & 0.04 & 20.27 & 0.04\\
HB33 &  182.5668 & 18.5771 & 21.60 & 0.03 & 21.29 & 0.03 & 20.59 & 0.04\\
HB34 &  182.5318 & 18.5426 & 20.99 & 0.03 & 21.05 & 0.04 & 20.07 & 0.03\\
HB35 &  182.5307 & 18.5430 & 21.96 & 0.04 & 21.55 & 0.05 & - & - \\
HB36 &  182.5292 & 18.5481 & 21.69 & 0.04 & 21.45 & 0.06 & - & - \\
HB37 &  182.5123 & 18.6111 & 21.48 & 0.05 & 21.68 & 0.05 & - & - \\
BSS01 &  182.5633 & 18.5480 & 21.80 & 0.05 & 21.81 & 0.05 & - & - \\
\hline\end{tabular} }

\end{table*}

\end{document}